\documentclass[conference]{IEEEtran}
\IEEEoverridecommandlockouts
\usepackage{cite}
\usepackage{amsmath,amssymb,amsfonts}
\usepackage{algorithmic}
\usepackage{graphicx}
\usepackage{textcomp}
\usepackage{orcidlink}
\usepackage{xcolor}
\usepackage{graphicx}
\usepackage{hyperref}
\usepackage{float}
\usepackage{stfloats}
\usepackage{subcaption}
\def\BibTeX{{\rm B\kern-.05em{\sc i\kern-.025em b}\kern-.08em
    T\kern-.1667em\lower.7ex\hbox{E}\kern-.125emX}}
\begin{document}

\title{Generative Quantum-inspired Kolmogorov-Arnold Eigensolver
\thanks{The views expressed in this article are those of the authors and do not represent the views of Wells Fargo. This article is for informational purposes only. Nothing contained in this article should be construed as investment advice. Wells Fargo makes no express or implied warranties and expressly disclaims all legal, tax, and accounting implications related to this article.}
}

\author{
\IEEEauthorblockN{
    Yu-Cheng Lin$^{1}$\IEEEauthorrefmark{1}\orcidlink{0009-0009-1911-9566}, 
    Yu-Chao Hsu$^{2,3}$\IEEEauthorrefmark{1}\orcidlink{0009-0004-7221-3854}, 
    I-Shan Tsai$^{4}$\orcidlink{0009-0005-5654-1986},\\
    Chun-Hua Lin$^{2,5}$\orcidlink{0009-0002-4383-0453},
    Kuo-Chung Peng$^{2,5}$\orcidlink{0009-0001-8342-2481},
    Jiun-Cheng Jiang$^{5,6,7}$\orcidlink{0009-0005-1134-4962},
    Yun-Yuan Wang$^{6}$\orcidlink{0009-0001-0323-3382},\\
    Tzung-Chi Huang$^{6}$,
    Tai-Yue Li$^{2}$\orcidlink{0000-0002-1993-1863},
    Kuan-Cheng Chen$^{8,9}$\IEEEauthorrefmark{3}\orcidlink{0000-0002-6575-7034},
    Samuel Yen-Chi Chen$^{10}$\IEEEauthorrefmark{4}\orcidlink{0000-0003-0114-4826},
    Nan-Yow Chen$^{2}$\IEEEauthorrefmark{5}\orcidlink{0000-0001-8139-6809}
}
\IEEEauthorblockA{
    $^1$Department of Electrophysics, National Yang Ming Chiao Tung University, Hsinchu, Taiwan.\\
    $^2$National Center for High-Performance Computing, National Institutes of Applied Research, Hsinchu, Taiwan\\
    $^3$Cross College Elite Program, National Cheng Kung University, Tainan, Taiwan \\
    $^4$Department of Mathematics, University of California, San Diego, San Diego, California, USA \\
    $^5$ Department of Physics and Center for Theoretical Physics, National Taiwan University, Taipei, Taiwan \\
    $^6$NVIDIA AI Technology Center, NVIDIA Corp., Taipei, Taiwan \\
    $^7$Center for Quantum Science and Engineering, National Taiwan University, Taipei, Taiwan\\
    $^8$Department of Electrical and Electronic Engineering, Imperial College London, London, UK\\
    $^9$Centre for Quantum Engineering, Science and Technology, Imperial College London, London, UK\\
    $^{10}$Wells Fargo, New York, NY, USA\\
}
\IEEEauthorblockA{\IEEEauthorrefmark{1}These authors contributed equally to this work.}

\IEEEauthorblockA{Emails: 
\IEEEauthorrefmark{3}\href{mailto:kuan-cheng.chen17@imperial.ac.uk}{kuan-cheng.chen17@imperial.ac.uk}, 
\IEEEauthorrefmark{4}\href{mailto:ycchen1989@ieee.org}{ycchen1989@ieee.org}, 
\IEEEauthorrefmark{5}\href{mailto:nanyow@nchc.narl.org.tw}{nanyow@nchc.narl.org.tw}
}
\vspace{-20pt}
}
\maketitle

\begin{abstract}
High-performance computing (HPC) is increasingly important for scalable quantum chemistry workflows in which classical generative models, quantum circuit simulation, and selected configuration interaction postprocessing are tightly coupled at scale.
This paper presents the generative quantum-inspired Kolmogorov--Arnold eigensolver (GQKAE), a parameter-efficient extension of the generative quantum eigensolver (GQE) for quantum chemistry problems.
Existing GPT-style generative eigensolvers formulate circuit construction as autoregressive sequence generation; however, their dense nonlinear feed-forward network (FFN) components introduce substantial parameter and memory overhead as molecular active spaces, operator pools, and circuit sequence lengths increase.
To address this limitation, GQKAE replaces these parameter-heavy FFN with hybrid quantum-inspired Kolmogorov--Arnold network (HQKAN) modules, resulting in a compact HQKANsformer backbone.
The proposed architecture preserves the autoregressive operator-selection mechanism and the quantum-selected configuration interaction (QSCI) evaluation pipeline, while incorporating expressive nonlinear mappings through single-qubit DatA Re-Uploading ActivatioN (DARUAN).
In contrast to variational quantum eigensolvers, which optimize continuous parameters within a fixed ansatz, GQKAE learns a discrete policy over shallow, task-specific circuit constructions and is trained using a QSCI-based reward signal.
Numerical benchmarks are performed on molecular systems involving bond dissociation, conformational variation, and intermolecular interactions, including H$_4$, N$_2$, LiH, C$_2$H$_6$, H$_2$O, and the H$_2$O dimer.
Across these benchmarks, GQKAE achieves chemical accuracy comparable to the GPT-based architecture employed in GQE, while reducing trainable parameters and memory by approximately 66\% and achieving the notable wall-time speedup.
For strongly correlated systems such as N$_2$ and LiH, GQKAE also improves convergence behavior and final energy errors.
These results demonstrate that quantum-inspired Kolmogorov--Arnold network (QKAN) can reduce classical-side overhead while preserving circuit-generation quality, providing a scalable approach for HPC-quantum co-design on near-term quantum platforms.
\end{abstract}


\section{Introduction}

High-performance computing (HPC) is becoming an essential layer of scalable quantum chemistry, where electronic-structure preprocessing, quantum-circuit simulation, model training, and selected configuration-interaction postprocessing must be coupled efficiently~\cite{liu2024quantumcentric,hicks2024massively,robledo2025chemistry}. This requirement is particularly important for near-term quantum algorithms because deep circuits, repeated sampling, and classical diagonalization can dominate end-to-end cost as active spaces, operator pools, and sequence lengths grow~\cite{huggins2021efficient,kanno2023quantum,kemmoku2026gqsci}. Thus, practical eigensolver design must consider not only quantum accuracy and circuit depth, but also the parameter and memory footprint of the classical components.

Estimating ground-state properties of many-body quantum systems is a central task in quantum chemistry and condensed-matter physics, with implications for molecular modeling and materials design~\cite{babbush2025grandchallenge}. Quantum machine learning (QML) provides a framework for learning and optimization with quantum models, including quantum neural networks and kernel methods~\cite{mitarai2018qcl,thanasilp2024exponential,jerbi2023quantum,hsu2025adaptiveexcitation,liu2024quantum,liu2025quantum,hsu2025federated,sim2019expressibility,hsu2025quantum,hsu2024kernel,tsai2025learning,lin2025meta,huang2026noise,romero2017quantum,Havlicek2019}. Within this landscape, variational quantum algorithms estimate ground-state energies by optimizing parameterized trial states. The variational quantum eigensolver (VQE)~\cite{peruzzo2014variational,mcclean2016theory,mcardle2020quantum,o2016scalable} has therefore become a leading approach in the noisy intermediate-scale quantum (NISQ) era~\cite{preskill2018NISQ}. However, VQE performance depends strongly on ansatz design and continuous-parameter optimization, and can be limited by expressivity constraints, noise, and barren plateaus~\cite{wang2021noise,Larocca2025barrenplateaus,mcclean2018barren,cerezo2022challenges}.

The generative quantum eigensolver (GQE)~\cite{nakaji2024gqe,minami2025generative,tyagin2025qaoa,keithley2026auger} addresses these limitations by reformulating ground-state preparation as discrete circuit generation. A classical generative model learns a probability distribution over sequences of unitaries selected from an operator pool, and sampled sequences define candidate circuits whose energies guide training. Because the trainable parameters reside in the classical generator rather than in the quantum circuit, GQE avoids direct optimization of parameterized quantum circuits. A representative implementation, the generative pretrained transformer-based eigensolver, uses a decoder-only transformer~\cite{radford2019language,vaswani2017attention} as the autoregressive circuit generator. Recent extensions combine GQE with quantum-selected configuration interaction (QSCI)~\cite{kanno2023quantum,nakagawa2024adapt,mikkelsen2025quantum,sugisaki2025hamiltonian,reinholdt2025critical,kaliakin2025implicit,robledo2025chemistry,kemmoku2026gqsci}, where sampled bitstrings define a truncated determinant subspace for classical Hamiltonian diagonalization.

Despite these advantages, the classical backbone of GPT-style GQE can become a scalability bottleneck. In transformers, dense position-wise nonlinear transformations account for a substantial fraction of the model parameters~\cite{geva2021transformer,vaswani2017attention,radford2019language}. As the operator vocabulary and generated sequence length increase, this parameter overhead raises memory use and training cost, directly affecting HPC-enabled quantum workflows. A compact generator that preserves circuit quality while reducing classical-side overhead is therefore desirable.

The recently proposed quantum-inspired Kolmogorov--Arnold network (QKAN)~\cite{jiang2025qkan} offers a promising route to such compression. QKAN realizes learnable edge functions through DatA Re-Uploading ActivatioN (DARUAN) modules~\cite{jiang2025qkan}, inspired by single-qubit data re-uploading circuits~\cite{perez2020data} whose Fourier spectrum expands with repeated encoding~\cite{Schuld_2021}. Its hybrid form, HQKAN, places a QKAN latent processor between a classical encoder and decoder, providing an expressive nonlinear transformation with reduced parameter cost.

In this work, we propose the generative quantum-inspired Kolmogorov--Arnold eigensolver (GQKAE), an extension of the general GQE framework that integrates HQKAN into its core generative backbone, yielding an HQKANsformer architecture that preserves autoregressive operator selection and the QSCI evaluation pipeline~\cite{kemmoku2026gqsci} while replacing parameter-heavy nonlinear transformations with compact QKAN-based modules. We apply GQKAE to H$_4$, N$_2$, LiH, C$_2$H$_6$, H$_2$O, and the H$_2$O dimer, trained with Group Relative Policy Optimization (GRPO)~\cite{shao2024deepseekmath} and simulated with CUDA-Q~\cite{kim2023cuda}.

Our contributions are summarized as follows:
\begin{enumerate}
    \item We introduce GQKAE, an HQKANsformer-based generative eigensolver for QSCI-guided molecular ground-state estimation.

    \item We show that GQKAE achieves chemical accuracy with quantum resource costs comparable to GQE across bond dissociation, conformational variation, and intermolecular interaction benchmarks.

    \item We demonstrate an approximately $66\%$ reduction in trainable parameters and parameter memory, along with notable speedups in wall time, effectively reducing classical-side overhead in scalable HPC-quantum co-design for quantum chemistry.
\end{enumerate}

The remainder of this paper is organized as follows. Sec.~\ref{sec:Related} reviews related work. Sec.~\ref{sec:Pre} introduces GQE, QSCI, and QKAN. Sec.~\ref{sec:GQKAE} presents GQKAE. Sec.~\ref{sec:Result} reports numerical results, and Sec.~\ref{sec:Discussion} concludes the paper.

\section{Related work}\label{sec:Related}

\subsection{Classical and quantum methods for quantum chemistry}
Classical electronic-structure methods span a hierarchy of accuracy and cost. 
Mean-field approaches such as Hartree--Fock (HF)~\cite{fukutome1981unrestricted} provide efficient baseline approximations. 
Perturbative and coupled-cluster methods including Coupled-Cluster Singles, Doubles and Triples (CCSD and CCSD(T))~\cite{kallay2001higher,helgaker2013molecular} achieve high accuracy in weakly correlated systems.
Strongly correlated methods such as Full Configuration Interaction (FCI), Complete Active Space Configuration Interaction (CASSCI), selected Configuration Interaction (SCI), and DMRG~\cite{holmes2016heat,sharma2017semistochastic,chan2011density,baiardi2020density} are systematically improvable but incur exponential cost, which restricts tractable active spaces~\cite{levine2020casscf,gao2024distributed,shayit2025numerically}.

VQE follows a similar progression. 
Fixed-ansatz approaches such as hardware-efficient circuits and unitary coupled-cluster~\cite{peruzzo2014variational,anand2022quantum,romero2019strategies,guo2024experimental,kandala2017hardware,motta2023bridging} embed parameters directly in quantum circuits. 
Adaptive methods including ADAPT-VQE and layer-wise constructions~\cite{grimsley2019adaptive,tang2021qubit,liu2022layer,weaving2025contextual,matousek2024variational} iteratively grow circuit expressivity.
Optimization-limited regimes arise due to barren plateaus~\cite{mcclean2018barren,cerezo2022challenges}, motivating approaches that shift learning toward classical components.

\subsection{Machine-learning-based quantum circuit design and KAN architectures}
Machine learning offers a complementary route to quantum circuit design and quantum architecture search (QAS), broadly grouped into three families of methods~\cite{martyniuk2024quantum}.
Reinforcement-learning (RL) approaches train an agent to construct circuits gate-by-gate using measurement-driven rollouts~\cite{ostaszewski2021reinforcement,fosel2021quantum,kuo2021quantum,patel2024curriculum}, while differentiable and one-shot QAS methods relax the discrete gate selection into a continuous domain or share parameters across a supernet to enable efficient gradient-based search~\cite{zhang2022differentiable,wu2023quantumdarts,wang2022quantumnas}.
In contrast, generative approaches such as GQE formulate circuit synthesis as autoregressive sequence modeling, avoiding intermediate measurements and shifting the optimization burden onto a classical generator~\cite{nakaji2024gqe,radford2019language,vaswani2017attention,keithley2026auger,holden2026spingqe}.
And hybrid approaches combine sampled subspaces with classicaldiagonalization~\cite{kanno2023quantum,kemmoku2026gqsci}.

KAN~\cite{liu2025kan} replaces fixed node activations with learnable univariate functions, enabling expressive function approximation with favorable parameter efficiency.
KAN modules have been adopted in quantum chemistry and quantum circuit design, including molecular property prediction~\cite{li2025kolmogorov} and RL-based architecture search for VQE~\cite{kundu2024kanqas}.
At larger scales, KAN-based transformers such as KAT~\cite{yang2025kolmogorovarnold} improve trainability via grouped edge functions, but scalability remains limited by the growth of trainable activations with layer dimensions. 
We therefore adopt the quantum-inspired KAN with DARUAN activations along with HQKAN~\cite{jiang2025qkan}, which preserves expressive edge-based parameterization while demonstrating scalability to large language models (LLMs).

\section{Preliminaries}\label{sec:Pre}

\subsection{Generative Quantum Eigensolver}

The GQE~\cite{nakaji2024gqe} formulates quantum ground-state search as a circuit-generation problem for a given Hamiltonian $\hat{H}$. 
Unlike the VQE, which optimizes continuous variational parameters within predetermined circuit family, GQE employs a trainable generative model to construct candidate quantum circuits by sequentially selecting operators from a predefined pool. Here, the ground-state search is recast as learning a probability distribution over circuit constructions.

To formalize this framework, we consider an operator pool derived from the unitary coupled-cluster singles and doubles (UCCSD) ansatz~\cite{taube2006new,cooper2010benchmark,lee2019generalized,evangelista2019exact,bauman2019quantum},
\begin{equation}
\mathcal{G}=\{\hat{U}_j\}_{j=1}^{|\mathcal{G}|},
\end{equation}
where each $\hat{U}_j$ denotes a unitary operator associated with an excitation term in the UCCSD pool, and $|\mathcal{G}|$ is the total number of operators. For a circuit of length $L$, a candidate circuit is specified by an operator-index sequence
$\boldsymbol{j}=(j_1,j_2,\dots,j_L),$
which determines an ordered composition of operators drawn from $\mathcal{G}$. The corresponding circuit can therefore be written as
$\hat{U}_{L}(\boldsymbol{j})=\hat{U}_{j_|\mathcal{G}|}\cdots \hat{U}_{j_2}\hat{U}_{j_1}.$
Acting on an initial reference state $|\psi_0\rangle$, this circuit prepares the trial state
$|\psi(\boldsymbol{j})\rangle=\hat{U}_{N}(\boldsymbol{j})|\psi_0\rangle.$
The quality of the sampled circuit is then evaluated through the expectation value of the target Hamiltonian,
\begin{equation}
E(\boldsymbol{j})=\langle \psi(\boldsymbol{j})|\hat{H}|\psi(\boldsymbol{j})\rangle.
\end{equation}
Accordingly, the objective of GQE is to train the generative model such that operator sequences assigned higher probability increasingly correspond to circuits that prepare lower-energy states.

Let $\boldsymbol{\theta}$ denote the trainable parameters of the generative model, and let $p_L(\boldsymbol{\theta},\boldsymbol{j})$ denote the induced probability distribution over operator sequences of length $L$. Since the sequence $\boldsymbol{j}$ is generated in an ordered manner, this distribution admits an autoregressive decomposition of the form
\begin{equation}
p_L(\boldsymbol{\theta},\boldsymbol{j})
=
\prod_{k=1}^{L}
p\bigl(j_k \mid \boldsymbol{j}_{<k};\boldsymbol{\theta}\bigr),
\label{eq:p_n}
\end{equation}
where $\boldsymbol{j}_{<k}=(j_1,\dots,j_{k-1})$ denotes the previously selected operators. This factorization makes GQE naturally compatible with autoregressive generative models, which learn the conditional probability of each operator given its preceding context. In this work, these conditional distributions are parameterized by a GPT-2-based architecture as our baseline, enabling candidate quantum circuits to be generated token by token over the predefined operator pool.

\subsection{Quantum-selected configuration interaction}\label{sec:QSCI}

Given a trial state $|\psi(\boldsymbol{j})\rangle$ prepared by the generated circuit, we adopt the QSCI procedure~\cite{kemmoku2026gqsci} to evaluate its quality. Instead of directly estimating the full expectation value of the Hamiltonian, QSCI constructs a truncated subspace from measurement outcomes and performs a classical diagonalization within this subspace.

Specifically, let
\begin{equation}
|\psi(\boldsymbol{j})\rangle
=
\sum_{x} a_x(\boldsymbol{j}) \, |x\rangle,
\end{equation}
where $\{|x\rangle\}$ denotes the computational basis corresponding to Slater determinants. By performing repeated measurements in this basis, we obtain a set of sampled bitstrings, which define a subset of determinants
$\mathcal{D}(\boldsymbol{j})
=
\{\, x \;\text{sampled from}\; |\psi(\boldsymbol{j})\rangle \,\}.$

To control the computational cost, we retain at most $d_{\max}$ determinants according to their sampling frequency, resulting in a truncated determinant set $\mathcal{D}_{d_{\max}}(\boldsymbol{j})$. These determinants span a subspace
$\mathcal{S}(\boldsymbol{j})
=
\mathrm{span}\{\, |x\rangle : x \in \mathcal{D}_{d_{\max}}(\boldsymbol{j}) \,\}.$

Within this subspace, we construct the projected Hamiltonian matrix
$H_{mn}(\boldsymbol{j}) = \langle x_m | \hat{H} | x_n \rangle$,
where $x_m, x_n \in \mathcal{D}_{d_{\max}}(\boldsymbol{j})$,
and obtain the QSCI energy by solving the corresponding eigenvalue problem,
\begin{equation}
E_{\mathrm{QSCI}}(\boldsymbol{j})
=
\min_{\substack{|\phi\rangle \in \mathcal{S}(\boldsymbol{j}) \\ \langle \phi | \phi \rangle = 1}}
\langle \phi | \hat{H} | \phi \rangle.
\end{equation}

In this work, the QSCI energy serves as the evaluation signal for the generated circuit. In particular, we define the reward associated with a sampled operator sequence $\boldsymbol{j}$ as
\begin{equation}
r(\boldsymbol{j}) = - E_{\mathrm{QSCI}}(\boldsymbol{j}),
\label{eq:qsci_reward}
\end{equation}
such that circuits leading to lower subspace energies are assigned higher rewards. This formulation establishes a direct connection between the generative model and the QSCI-based evaluation, forming the core optimization loop of the GQE framework.

\subsection{Quantum-inspired Kolmogorov--Arnold Network}

The QKAN follows the structural principle of Kolmogorov--Arnold networks, in which multivariate mappings are constructed through learnable univariate transformations associated with edges rather than fixed node-wise activation functions. In the present work, these edge-wise nonlinear functions are implemented using the DARUAN, a quantum-inspired variational activation module built from a single-qubit data re-uploading circuit. Through repeated data encoding and trainable circuit parameters, DARUAN provides a compact and expressive family of learnable nonlinear mappings. 

Let
$\mathbf{x}_{l}=\left(x_{l,1},x_{l,2},\dots,x_{l,n_l}\right)$
denote the input vector to the $l$-th layer, where $n_l$ is the corresponding layer width. For each output unit $j$ in the next layer, QKAN aggregates the transformed contributions from all input coordinates according to
\begin{equation}
x_{l+1,j}
=
\sum_{i=1}^{n_l}\phi_{l,j,i}(x_{l,i}),
\end{equation}
where $\phi_{l,j,i}(\cdot)$ denotes the learnable univariate function associated with the edge from input coordinate $i$ to output coordinate $j$.

In our construction, each edge function is realized by a DARUAN module. For an input scalar $x$, the corresponding activation is defined as
\begin{equation}
\phi(x;\boldsymbol{\vartheta})
=
\langle 0|
\hat{U}^{\dagger}(x;\boldsymbol{\vartheta})
\hat{M}
\hat{U}(x;\boldsymbol{\vartheta})
|0\rangle,
\end{equation}
where $\hat{U}(x;\boldsymbol{\vartheta})$ denotes a parameterized  data re-uploading circuit and $\hat{M}$ is the measurement observable. Through repeated data re-uploading, DARUAN induces a rich Fourier spectrum, enabling QKAN to represent highly nonlinear mappings with relatively few trainable parameters. In this way, DARUAN serves as a quantum-inspired variational activation function that maps a scalar input to a learnable nonlinear response while maintaining both expressivity and parameter efficiency.

Using DARUAN as the edge activation, a QKAN layer defines the mapping
$
\mathbf{x}_{l+1}=\Phi_l(\mathbf{x}_l),
$
where $\Phi_l$ collects all edge-wise transformations and summations in layer $l$. By stacking multiple such layers, the overall QKAN model realizes a hierarchical nonlinear map,
\begin{equation}
\mathcal{F}_{\mathrm{QKAN}}
=
\Phi_{L_{\mathrm{net}}-1}
\circ
\Phi_{L_{\mathrm{net}}-2}
\circ
\cdots
\circ
\Phi_{0}.
\label{eq:qkan}
\end{equation}

Compared with standard multilayer perceptrons, QKAN shifts the main source of representation from node-wise affine projections followed by fixed activations to adaptive edge-wise nonlinear operators. This design is particularly suitable for generative sequence modeling, where expressive yet parameter-efficient conditional transformations are desirable.

\subsection{HQKAN architecture}

\begin{figure}[t]
\centering
\includegraphics[width=\columnwidth]{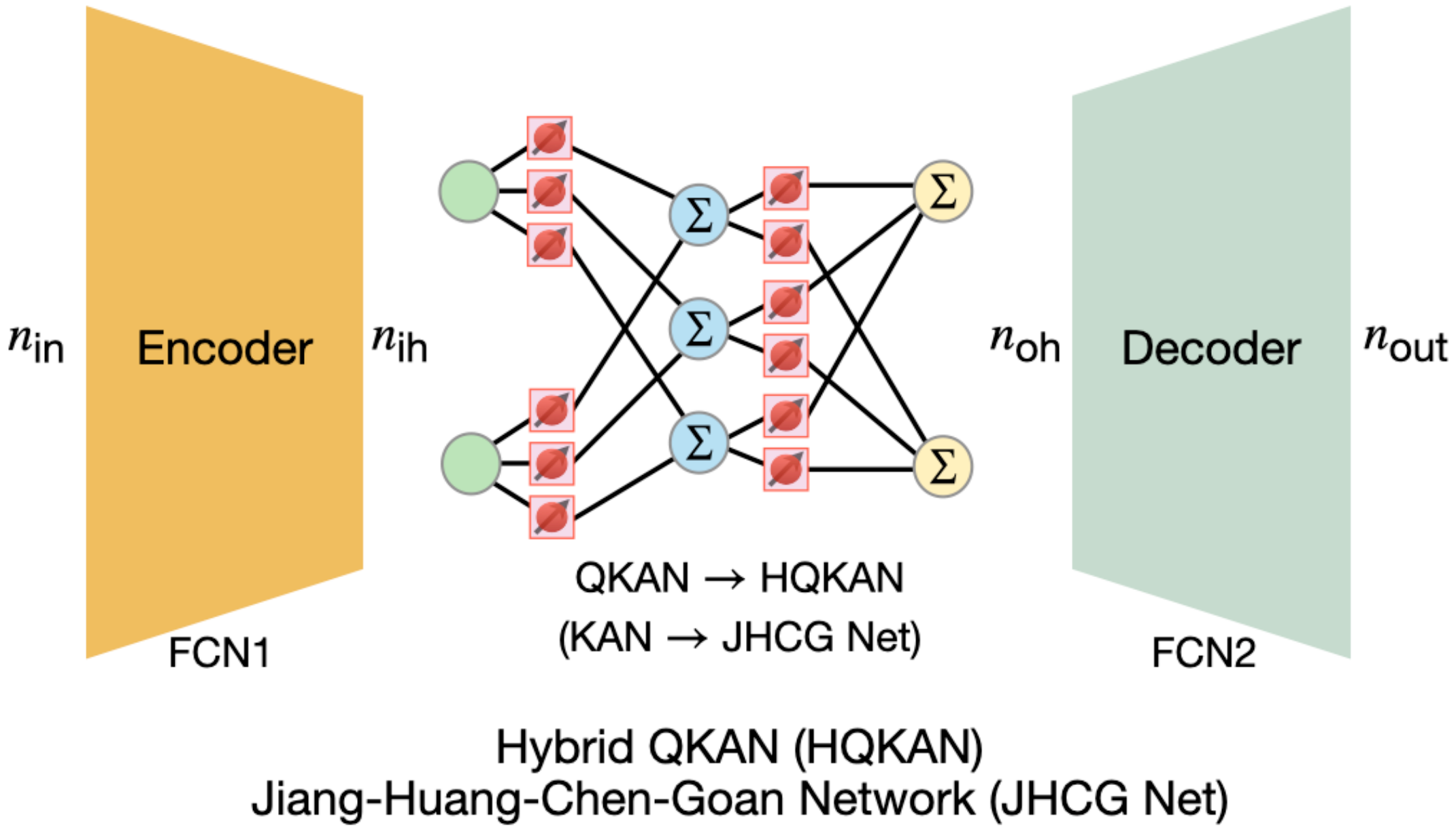}
\caption{
    \textbf{Architecture of the Jiang--Huang--Chen--Goan Network (JHCG Net)~\cite{jiang2025qkan}.}
     The JHCG Net uses an encoder--processor--decoder structure, where a QKAN-based latent processor forms the Hybrid QKAN (HQKAN) module and introduces quantum-inspired nonlinear transformations in the latent space.
    }
\label{fig:hqkan}
\vspace{-10pt}
\end{figure}
Ref.~\cite{jiang2025qkan,hsu2025qkan} further introduced the concept of HQKAN,
representing a fusion of classical and quantum-inspired neural computation.
As illustrated in Fig.~\ref{fig:hqkan}, the Jiang-Huang-Chen-Goan Network (JHCG Net) follows an encoder–processor–decoder design, in which the encoder first maps the input features into a compact latent representation. This latent representation is then processed by a KAN-based module, which provides flexible nonlinear function approximation through learnable univariate transformations. In the HQKAN setting, the latent KAN module is replaced or enhanced by QKANs, allowing the model to incorporate quantum-inspired nonlinear mappings within the latent feature space. Consequently, the decoder reconstructs the output representation from the processed latent features. Within our framework, HQKAN is adopted as the nonlinear latent processor to enhance representation flexibility in a parameter-efficient manner. Its integration into the proposed generative model will be specified in the following section.

\section{Method}\label{sec:GQKAE}
Our method follows the GQE-for-QSCI workflow~\cite{kemmoku2026gqsci}, in which an autoregressive generative model produces operator sequences that define candidate circuits, the resulting quantum states are measured to sample Slater determinants, and the Hamiltonian is classically diagonalized in the sampled subspace to evaluate circuit quality. Within this workflow, our goal is not to modify the QSCI post-processing pipeline, but to improve the generative backbone used for circuit construction. To this end, we replace the conventional feed-forward mapping in the original GPT-2-based generator with an HQKAN-architecture module~\cite{jiang2025qkan}. The resulting framework is referred to as the generative quantum-inspired Kolmogorov--Arnold eigensolver (GQKAE), and the corresponding HQKAN-enhanced transformer backbone is hereafter referred to as the \emph{HQKANsformer}.

\subsection{Generative Quantum-inspired Kolmogorov--Arnold
Eigensolver}

\begin{figure*}[!t]
    \centering
    \vspace{-10pt}
    \includegraphics[width=\textwidth]{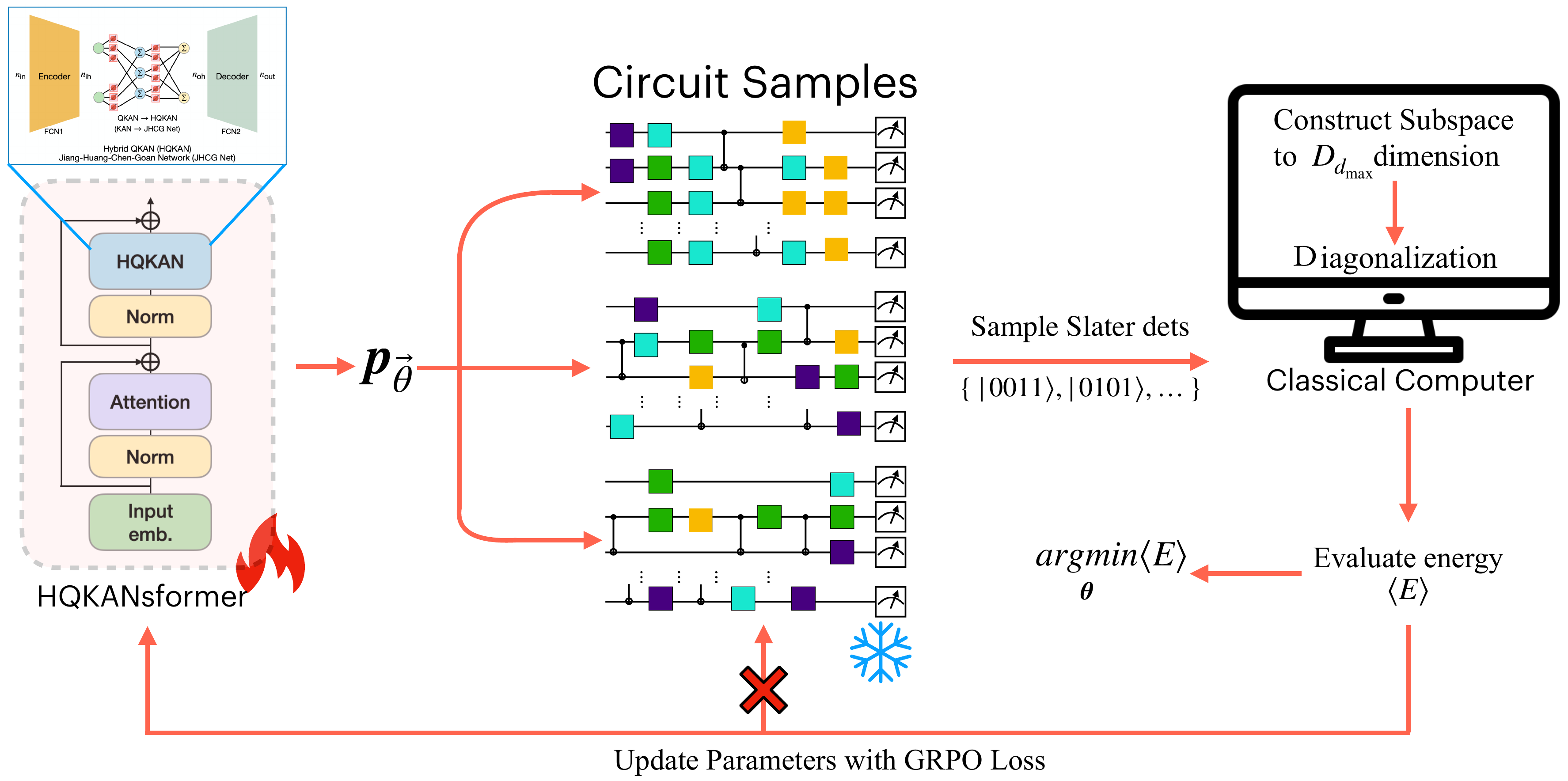}
    \caption{\textbf{Overview of the GQKAE framework.}
    The HQKANsformer defines an autoregressive distribution over operator sequences, which are used to construct candidate quantum circuits. The generated circuits prepare trial states that are measured to obtain sampled Slater determinants. These determinants span a truncated subspace of dimension $D_{d_{\max}}$, where the Hamiltonian is classically diagonalized to compute the QSCI energy. This energy is converted into a reward signal for updating the generative model via GRPO loss~\cite{shao2024deepseekmath}. The proposed method modifies only the generative backbone, while preserving the QSCI-based evaluation workflow.}
    \vspace{-15pt}
\end{figure*}

To reduce the number of trainable parameters in the GPT-2 backbone while preserving its autoregressive sequence-modeling capability, we replace the conventional FFN in each transformer block with an HQKAN module. 

In this way, the proposed GQKAE retains the standard GQE formulation at the level of circuit generation, while adopting a more parameter-efficient nonlinear transformation for modeling the conditional distribution over operator tokens.

More specifically, let $\mathbf{h}_{k}^{(l)} \in \mathbb{R}^{d}$ denote the hidden representation at sequence position $k$ in the $l$-th transformer block. In the standard GPT-2 architecture, the FFN is a two-layer nonlinear mapping of the form
\(W_2\sigma(W_1\mathbf{h}_{k}^{(l)}+\mathbf{b}_1)+\mathbf{b}_2\),
where $W_{1}\in\mathbb{R}^{d_{\mathrm{ff}}\times d}$ and $W_{2}\in\mathbb{R}^{d\times d_{\mathrm{ff}}}$ are learnable weight matrices, and $\sigma(\cdot)$ denotes the activation function. Since the intermediate dimension $d_{\mathrm{ff}}$ is typically much larger than $d$, the FFN contributes a substantial portion of the total parameter count.

In GQKAE, we replace this parameter-heavy feed-forward mapping by an HQKAN architecture transformation. Given the input hidden state $\mathbf{h}_{k}^{(l)}$, we first project it into a lower-dimensional latent space,
$
\mathbf{z}_{k}^{(l)}
=
W_{\mathrm{enc}}^{(l)}\mathbf{h}_{k}^{(l)}+\mathbf{b}_{\mathrm{enc}}^{(l)},
$
where $W_{\mathrm{enc}}^{(l)}\in\mathbb{R}^{d_z\times d}$ and $d_z<d$, with $d_z$ denoting the latent dimension of the HQKAN module.
\begin{equation}
\tilde{\mathbf{z}}_{k}^{(l)}
=
\mathcal{F}_{\mathrm{QKAN}}^{(l)}\!\left(\mathbf{z}_{k}^{(l)}\right),
\end{equation}
where $\mathcal{F}_{\mathrm{QKAN}}^{(l)}$ denotes the QKAN mapping in the $l$-th block, as defined in Eq.~\ref{eq:qkan}. Following the formulation introduced above, each component of $\tilde{\mathbf{z}}_{k}^{(l)}$ is obtained through edge-wise DARUAN activations,
\begin{equation}
\tilde{z}_{k,j}^{(l)}
=
\sum_{i=1}^{d_z}
\phi_{j,i}^{(l)}\!\left(z_{k,i}^{(l)}\right),
\qquad j=1,\dots,d_z,
\end{equation}
where each $\phi_{j,i}^{(l)}(\cdot)$ is implemented by a DARUAN module. The transformed latent vector is finally projected back to the original hidden dimension,
$
\mathbf{g}_{k}^{(l)}
=
W_{\mathrm{dec}}^{(l)}\tilde{\mathbf{z}}_{k}^{(l)}+\mathbf{b}_{\mathrm{dec}}^{(l)},
$
with $W_{\mathrm{dec}}^{(l)}\in\mathbb{R}^{d\times d_z}$.

Accordingly, the hidden-state update in the feed-forward part of the transformer block becomes
$
\mathbf{h}_{k}^{(l+1)}
=
\mathbf{h}_{k}^{(l)}
+
\mathbf{g}_{k}^{(l)},
$
up to the standard layer normalization and self-attention operations inherited from the GPT-style architecture. Therefore, the overall autoregressive backbone remains unchanged in structure, while its nonlinear feed-forward transformation is replaced by an HQKAN module whose latent-space processor is given by QKAN.

Given the final hidden representation $\mathbf{h}_{k}^{(L)}$ after $L$ transformer blocks, the logits over the operator vocabulary are computed as
\begin{equation}
\mathbf{o}_{k}
=
W_{\mathrm{out}}\mathbf{h}_{k}^{(L)}+\mathbf{b}_{\mathrm{out}},
\end{equation}
and the conditional probability of the next operator token is obtained as
$
p\!\left(j_k \mid \boldsymbol{j}_{<k};\boldsymbol{\theta}\right)
=
\mathrm{softmax}\!\left(\mathbf{o}_{k}\right).
$
Hence, the joint distribution over an operator sequence $\boldsymbol{j}=(j_1,\dots,j_N)$ remains autoregressive, as defined in Eq.~\ref{eq:p_n}, while the underlying conditional mapping is now parameterized by the HQKANsformer rather than the original GPT-2 backbone.

This replacement is further supported by the approximation efficiency of QKAN~\cite{jiang2025qkan}, which achieves a complexity scaling of $O(\log(1/\epsilon))$ for approximation error $\epsilon$. Thus, replacing the FFN with an encoder--QKAN--decoder HQKAN module with bottleneck dimension $d_z < d$ offers a parameter-efficient alternative to the standard GPT-2 feed-forward layer. In this sense, GQKAE extends GQE by modeling the generative distribution over circuit operators with an HQKANsformer backbone.

\subsection{GQKAE for QSCI}

With the HQKANsformer defining the autoregressive distribution over operator sequences, the remaining step is to specify how this policy is trained under the QSCI-based evaluation criterion. For each sampled sequence $\boldsymbol{j}$, the generated circuit prepares a trial state, from which measurement outcomes are collected to construct the truncated determinant subspace described in Sec.~\ref{sec:QSCI}. Classical diagonalization within this subspace yields the corresponding QSCI energy $E_{\mathrm{QSCI}}(\boldsymbol{j})$, which is converted into a
reward through Eq.~\ref{eq:qsci_reward}. In this manner, the proposed framework retains the original QSCI evaluation procedure and modifies only the generative policy used to produce candidate circuits.

To train the HQKANsformer-based generator, we optimize the autoregressive policy with a clipped GRPO objective~\cite{shao2024deepseekmath}, which updates the sequence model at the token level while using the quality of the full generated circuit as the learning signal. The corresponding loss is defined as
\begin{equation}
\begin{aligned}
\mathcal{L}_{\mathrm{GRPO}}(\boldsymbol{\theta})
&=
-\frac{1}{M}\sum_{m=1}^{M}\frac{1}{N}\sum_{t=1}^{N}
\min\!\left(
\rho_t^{(m)}(\boldsymbol{\theta})\,\hat{A}^{(m)}, \right. \\
&\quad \left.
\mathrm{clip}\bigl(
\rho_t^{(m)}(\boldsymbol{\theta}),
1-\epsilon,\,1+\epsilon
\bigr)\hat{A}^{(m)}
\right),
\end{aligned}
\end{equation}
where $\epsilon$ is the clipping parameter and
\begin{equation}
\rho_t^{(m)}(\boldsymbol{\theta})
=
\frac{
p_{\boldsymbol{\theta}}\!\left(
j_t^{(m)} \mid j_{<t}^{(m)}
\right)
}{
p_{\boldsymbol{\theta}_{\mathrm{old}}}\!\left(
j_t^{(m)} \mid j_{<t}^{(m)}
\right)
}
\end{equation}
denotes the token-wise importance ratio between the current policy and
the reference policy from which the samples were drawn.

The scalar signal $\hat{A}^{(m)}$ is constructed from the QSCI-based
evaluation of the entire sampled sequence. Concretely, for a batch of
sampled operator sequences
$\{\boldsymbol{j}^{(m)}\}_{m=1}^{M}$,
we first compute the reward of each sequence as
\(
r^{(m)} = r(\boldsymbol{j}^{(m)})
= -E_{\mathrm{QSCI}}(\boldsymbol{j}^{(m)}),
\)
so that sequences yielding lower QSCI energies receive larger rewards.
We then standardize the rewards within each batch to obtain a relative
advantage signal. Specifically, the normalized advantage is defined as
$
\hat{A}^{(m)}
=
\frac{r^{(m)}-\bar{r}}{\sigma_r},
$
where $\bar{r}$ and $\sigma_r$ denote the mean and standard deviation
of the rewards within the batch.

\section{Numerical Results}\label{sec:Result}
\subsection{Computational Details and Experiment Setup}
We evaluate the proposed framework across various molecular systems, comparing GQKAE against the original GQE baseline. All simulations ran on an NVIDIA HGX H200 system.
\begin{figure*}[!t]
    \vspace{-10pt}
    \centering
    
    \begin{subfigure}{0.32\textwidth}
        \centering
        \includegraphics[width=\linewidth]{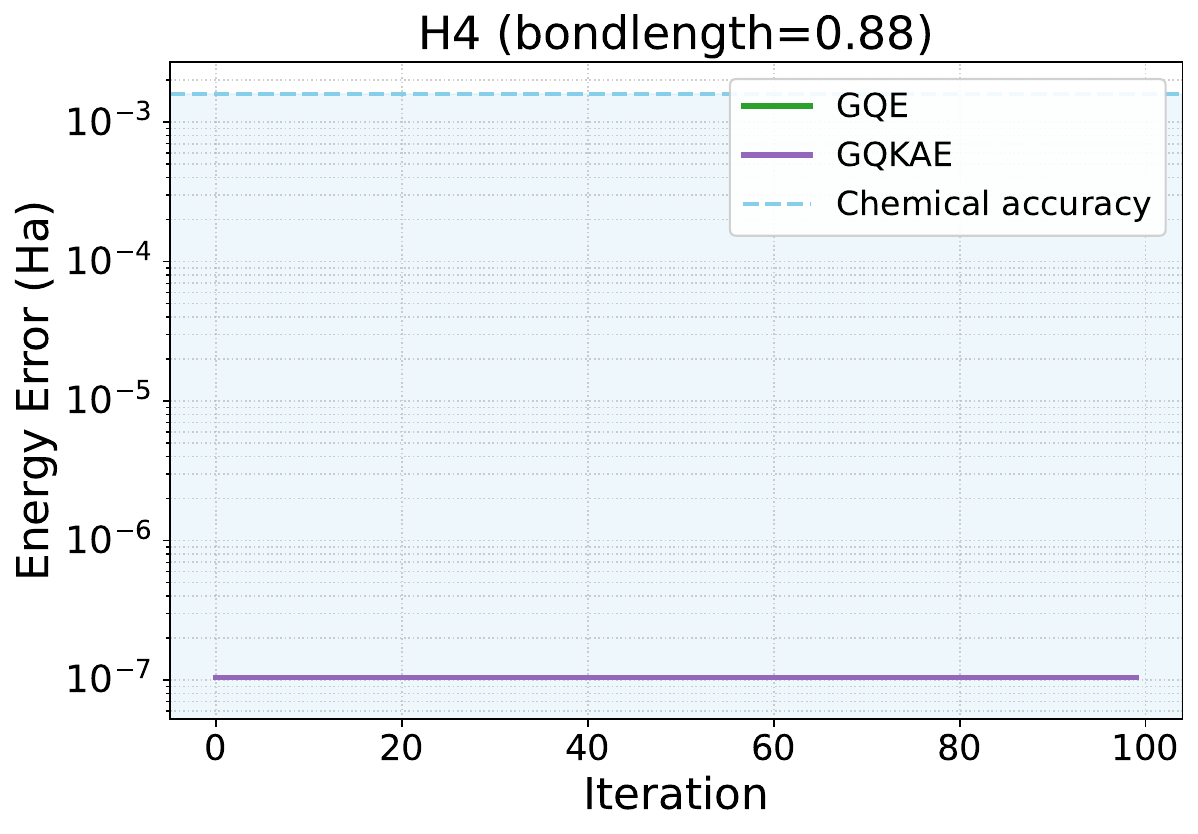}
        \caption{$\text{H}_4$ (8 qubits)}
        \label{fig:loss_h4}
    \end{subfigure}
    \hfill
    \begin{subfigure}{0.32\textwidth}
        \centering
        \includegraphics[width=\linewidth]{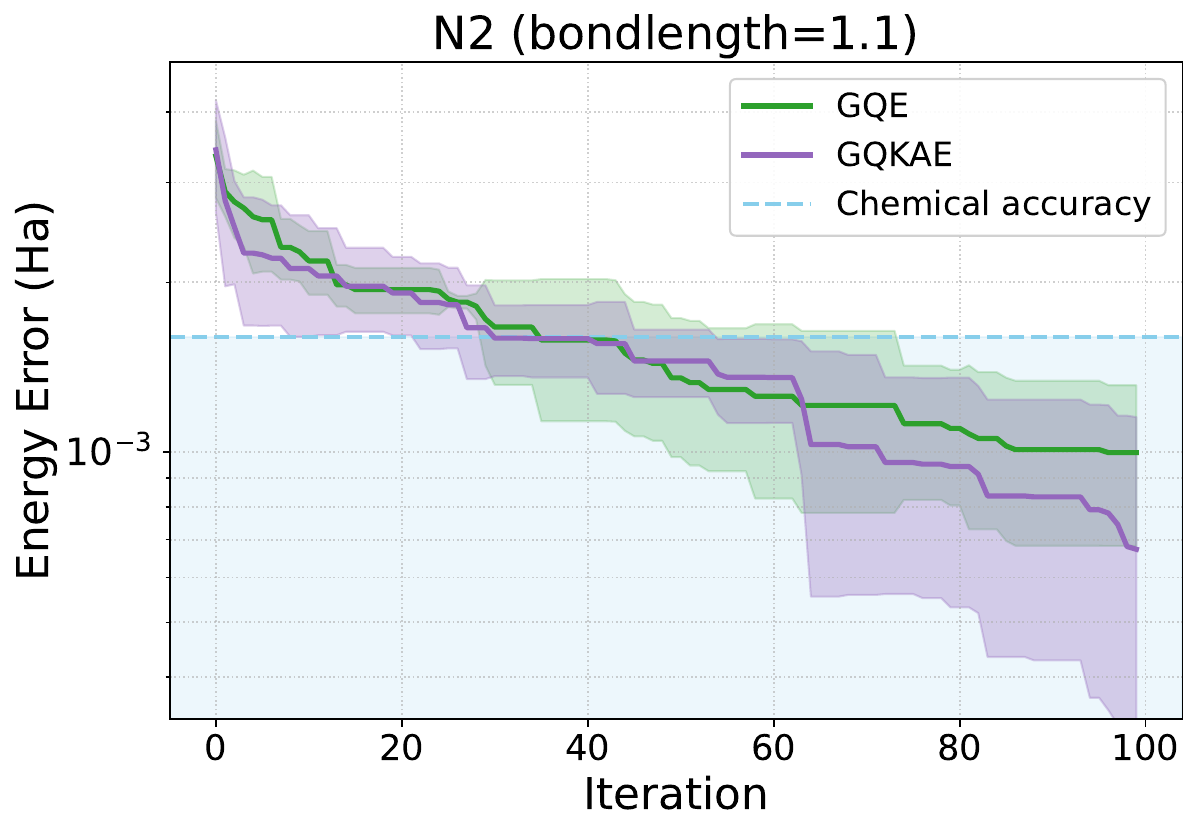}
        \caption{$\text{N}_2$ (16 qubits)}
        \label{fig:loss_n2}
    \end{subfigure}
    \hfill
    \begin{subfigure}{0.32\textwidth}
        \centering
        \includegraphics[width=\linewidth]{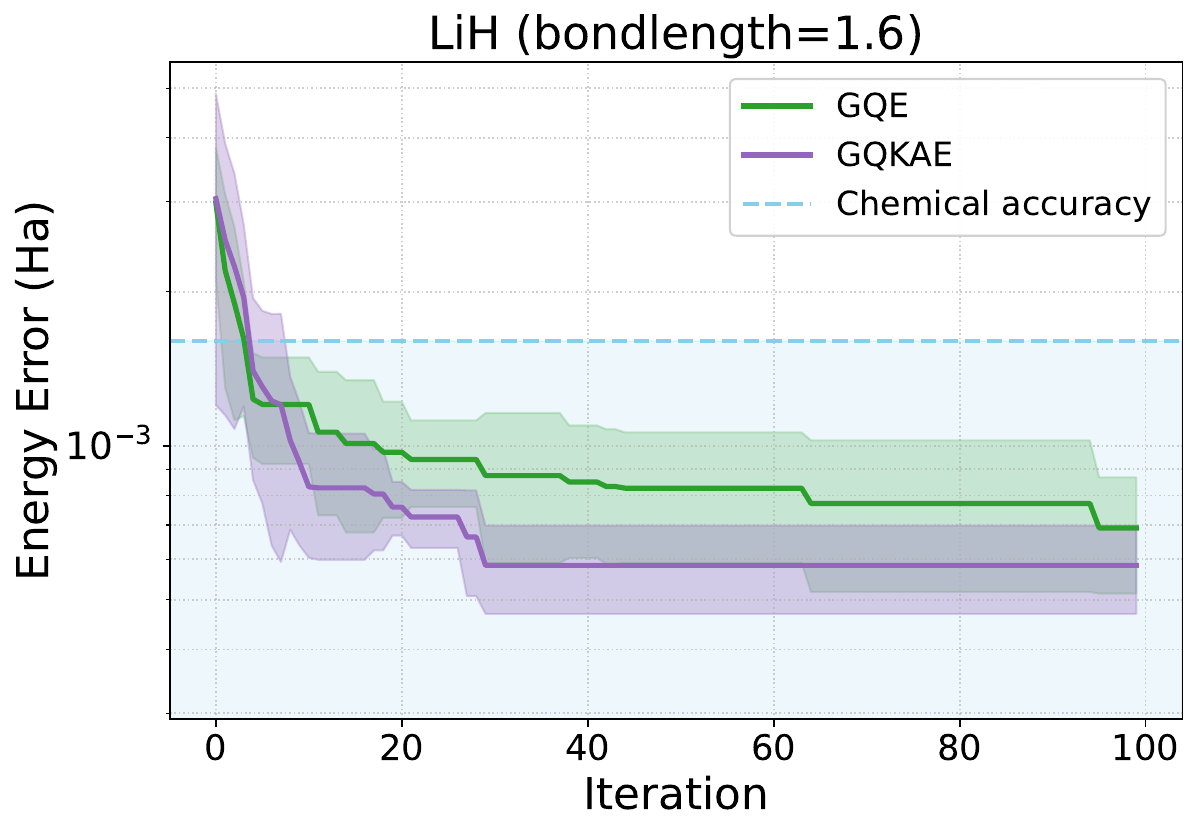}
        \caption{LiH (20 qubits)}
        \label{fig:loss_lih}
    \end{subfigure}

    \vspace{2pt}

    \begin{subfigure}{0.32\textwidth}
        \centering
        \includegraphics[width=\linewidth]{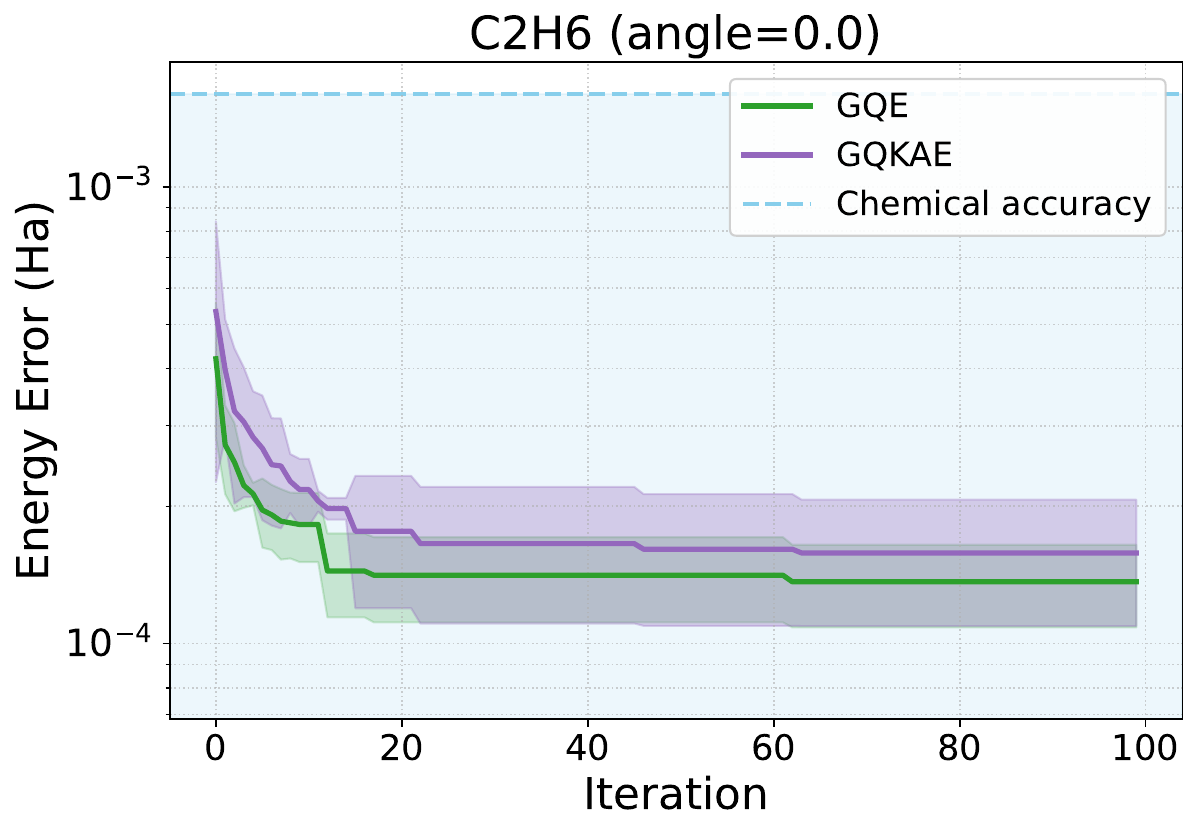}
        \caption{$\text{C}_2\text{H}_6$ (12 qubits)}
        \label{fig:loss_c2h6}
    \end{subfigure}
    \hfill
    \begin{subfigure}{0.32\textwidth}
        \centering
        \includegraphics[width=\linewidth]{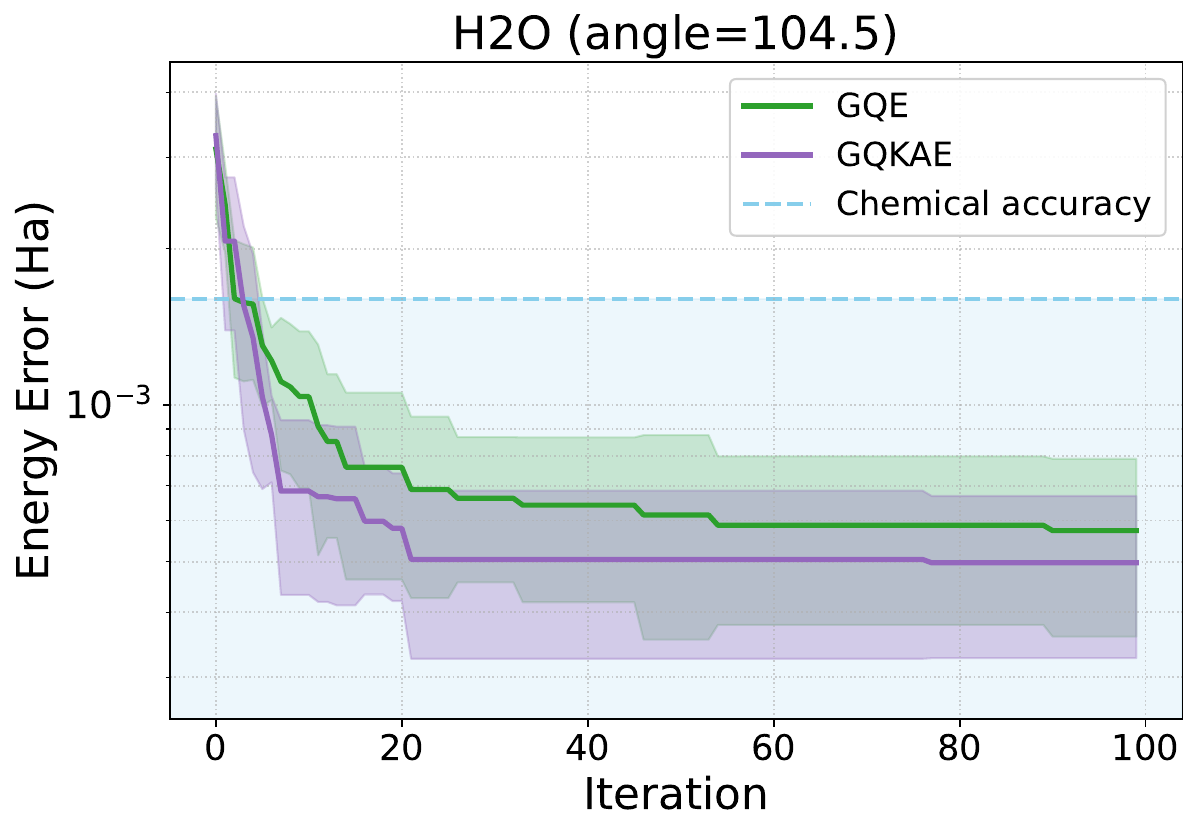}
        \caption{$\text{H}_2\text{O}$ (16 qubits)}
        \label{fig:loss_h2o}
    \end{subfigure}
    \hfill
    \begin{subfigure}{0.32\textwidth}
        \centering
        \includegraphics[width=\linewidth]{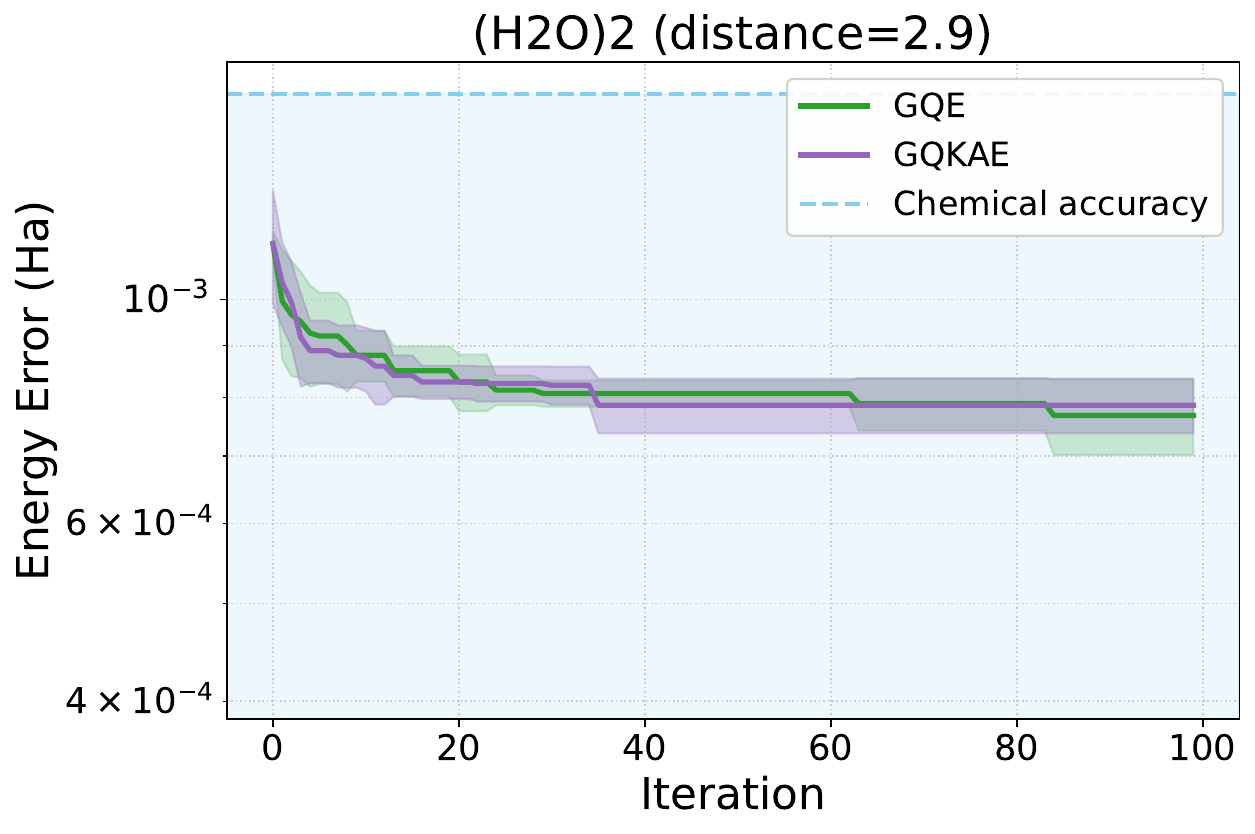}
        \caption{$\text{H}_2\text{O}$ dimer (16 qubits)}
        \label{fig:loss_h2o_dimer}
    \end{subfigure}

    \caption{Optimization history showing the best-so-far energy error relative to the CASCI baseline up to each iteration. Solid curves represent the mean values averaged over five independent optimization trials with different random seeds, while the shaded regions indicate one standard deviation. The horizontal dashed lines denote the chemical accuracy threshold (1.6 mHa).}
    \label{fig:loos_all}
    \vspace{-15pt}
\end{figure*}

Model performance was assessed by varying different physical properties, including bond lengths, bond angles, and intermolecular distances. The systems studied include H$_4$ $(4e, 4o)$, N$_2$ $(10e, 8o)$, and LiH $(4e, 10o)$ for bond dissociation analysis; C$_2$H$_6$ $(6e, 6o)$ and H$_2$O $(8e, 8o)$ for conformational angle variations; and the H$_2$O dimer $(8e, 8o)$ for intermolecular distance scanning. These configurations correspond to $8, 16, 20, 12, 16$, and $16$ qubits, respectively. The 6-31G~\cite{Hehre1972631g} was employed for H$_4$, LiH, and C$_2$H$_6$. The STO-3G basis~\cite{Hehre1969sto3g} was utilized for N$_2$, while the cc-pVDZ~\cite{dunning1989pVDZ} basis set was specifically adopted for both the H$_2$O monomer and the dimer. Molecular integrals and canonical basis using the restricted HF method via PySCF~\cite{sun2018pyscf}, which active spaces selected symmetrically around HOMO-LUMO gap.

For the policy network, we adopted a decoder-only GPT-2~\cite{radford2019language} architecture. To seamlessly integrate the quantum-inspired activations while controlling the parameter overhead, we embedded the QKAN layer within a highly compressed latent space. Specifically, the input was linearly projected from the embedding dimension down to a logarithmic latent dimension ($d_{\text{latent}}=12$) before applying the non-linear transformations using an open-source QKAN implementation adapted from~\cite{jiang2025qkan_github}\footnote{Available at \href{https://github.com/Jim137/qkan}{https://github.com/Jim137/qkan}.}. To further accelerate the framework, we adopted the efficient quantum-circuit solver, \texttt{FlashQKAN}, introduced in~\cite{jiang2025qkan_github}, which  leverages the \texttt{cuQuantum} library~\cite{10313722} to optimize the tensor-contraction path by representing each QKAN layer as a tensor network, while \texttt{cuTe DSL}~\cite{cecka2026cutelayoutrepresentationalgebra} is used for fused operator execution and block tiling to improve GPU throughput.

Following~\cite{kemmoku2026gqsci}\footnote{Available at \href{https://github.com/moken20/gqe-for-qsci}{https://github.com/moken20/gqe-for-qsci}.} , we sampled $M=10$ circuits per iteration with $N_{\text{shots}}=10^5$, training for $N_{\text{iter}}=100$ iterations. Policy parameters were optimized via GRPO~\cite{shao2024deepseekmath} and AdamW~\cite{Loshchilov2017AdamW} optimizer (learning rate $5\times 10^{-6}$, weight decay of $0.01$) with 30 policy updates per batch and a 1.2 repetition penalty. QSCI post-processing used a subspace dimension limit of $d_{\text{max}}=2000$ with symmetry completion. Simulations used CUDA-Q~\cite{kim2023cuda} with its natively optimized GPU acceleration, and PyCI~\cite{Richer2024PyCI}. Results average five independent trials.

To rigorously benchmark our generative models, we compared them against several classical and quantum methods. Classical references computed via PySCF include restricted HF energy, CCSD~\cite{Bartlett2007CCSD}, SCI, implemented as heat-bath CI~\cite{holmes2016heat,sharma2017semistochastic} via \texttt{pyscf.fci.SCI} with selection and CI-coefficient threshold of $5\times 10^{-4}$; and CASCI. Since the FCI dimension in our active space does not exceed $5\times10^3$, CASCI provides the exact ground state within this active space and serves as our variational reference. A model is considered to have achieved ``chemical accuracy" if its energy deviates from the CASCI energy by no more than 1.6 mHa ($\approx$ 1 kcal/mol).

For the quantum baseline, we employed the VQE with a UCCSD ansatz, simulated via CUDA-Q. VQE optimization used COBYLA (maximum iterations set to 5000) starting from a zero-initialized parameter state applied to the HF reference.

\subsection{Convergence Dynamics of Circuit Generation}

To demonstrate that the proposed GQKAE framework effectively optimizes the QSCI input state, we tracked the convergence behavior during the circuit generation process. As depicted in Fig.~\ref{fig:loos_all}, all target molecules were evaluated at their equilibrium geometries. We recorded the lowest energy error relative to the exact CASCI energy achieved up to each iteration step. 

Across the molecular systems, both GQE and GQKAE effectively guide the circuit structure toward lower energies, ultimately reaching the chemical accuracy threshold. For structurally simple systems such as H$_4$ (Fig.~\ref{fig:loss_h4}), the optimization landscape is relatively trivial. Consequently, both methods immediately discover highly accurate states, leading to nearly flat error curves that remain well below the chemical accuracy margin from the very first iterations.

The distinct advantages of the quantum-inspired activations emerge as the complexity of the electronic structure increases. In systems characterized by stronger correlation or complex orbital interactions, such as N$_2$, LiH, and H$_2$O (Fig.~\ref{fig:loss_n2},~\ref{fig:loss_lih}, and~\ref{fig:loss_h2o}), GQKAE achieves a noticeably lower final energy error than the standard GQE. Furthermore, particularly in the LiH and H$_2$O systems, GQKAE exhibits a steeper initial descent. This indicates that the enhanced non-linear expressivity provided by the QKAN layer allows the policy network to more rapidly identify optimal gate sequences before reaching a stable plateau.

For the remaining systems, including C$_2$H$_6$ and H$_2$O dimer, the two models exhibit highly comparable convergence trajectories. In the case of the C$_2$H$_6$, standard GQE locates a marginally lower energy minimum, though both models perform equivalently within the statistical variance.

\subsection{Results of Potential Energy Surfaces}
\begin{figure*}[t]
    \centering
    
    \begin{subfigure}{0.310\textwidth}
        \centering
        \includegraphics[width=\linewidth]{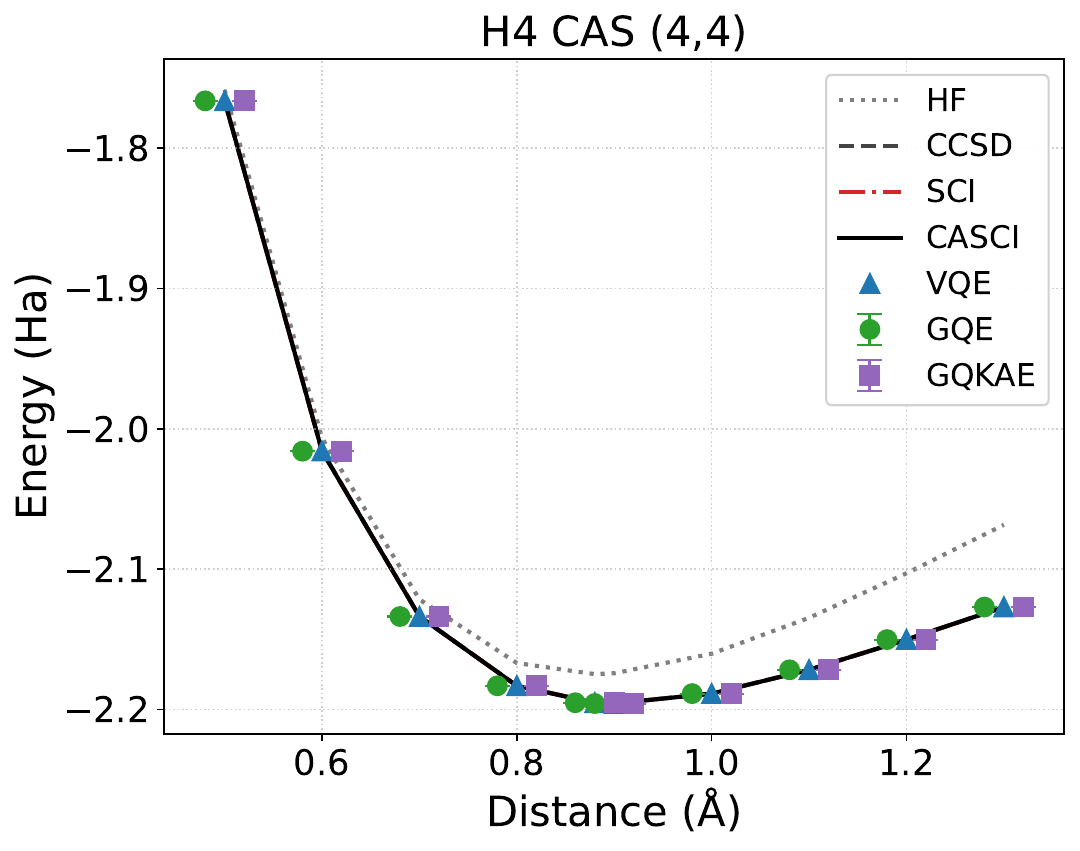}
        \caption{$\text{H}_4$ (8 qubits)}
        \label{fig:energy_h4}
    \end{subfigure}
    \hfill
    \begin{subfigure}{0.32\textwidth}
        \centering
        \includegraphics[width=\linewidth]{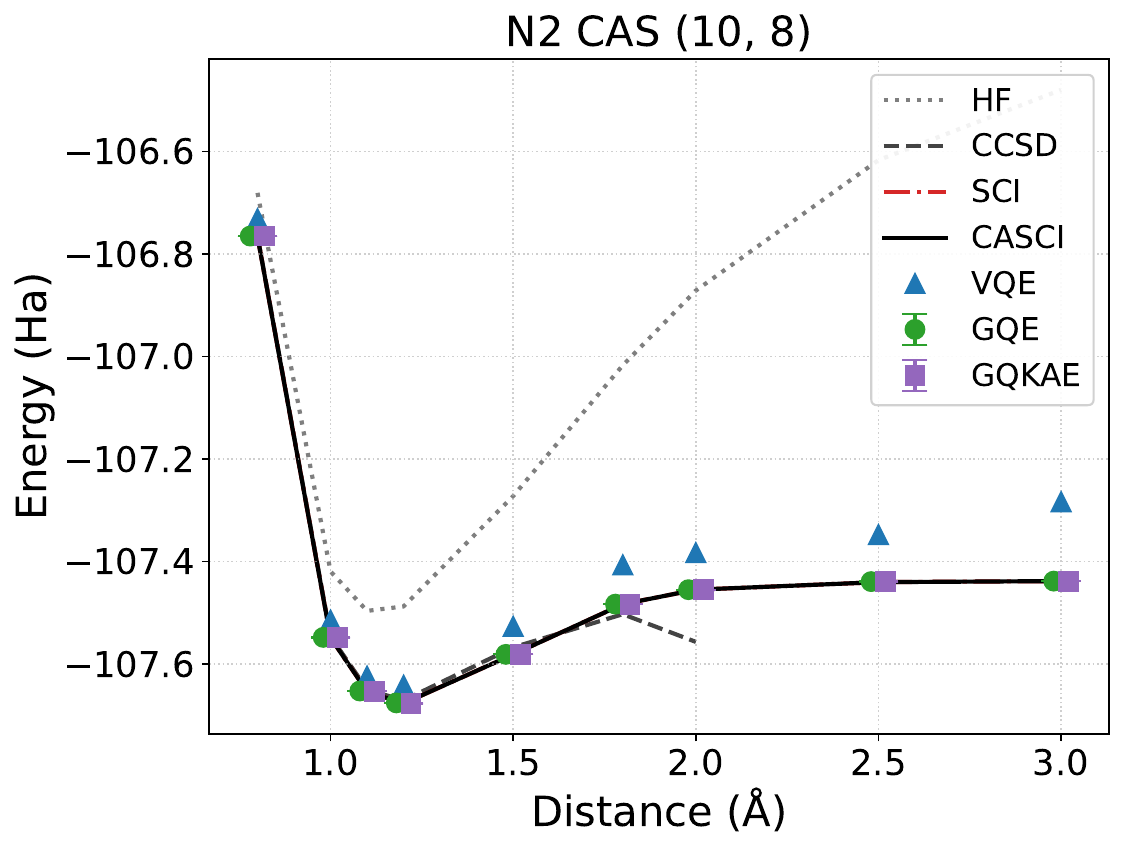}
        \caption{$\text{N}_2$ (16 qubits)}
        \label{fig:energy_n2}
    \end{subfigure}
    \hfill
    \begin{subfigure}{0.32\textwidth}
        \centering
        \includegraphics[width=\linewidth]{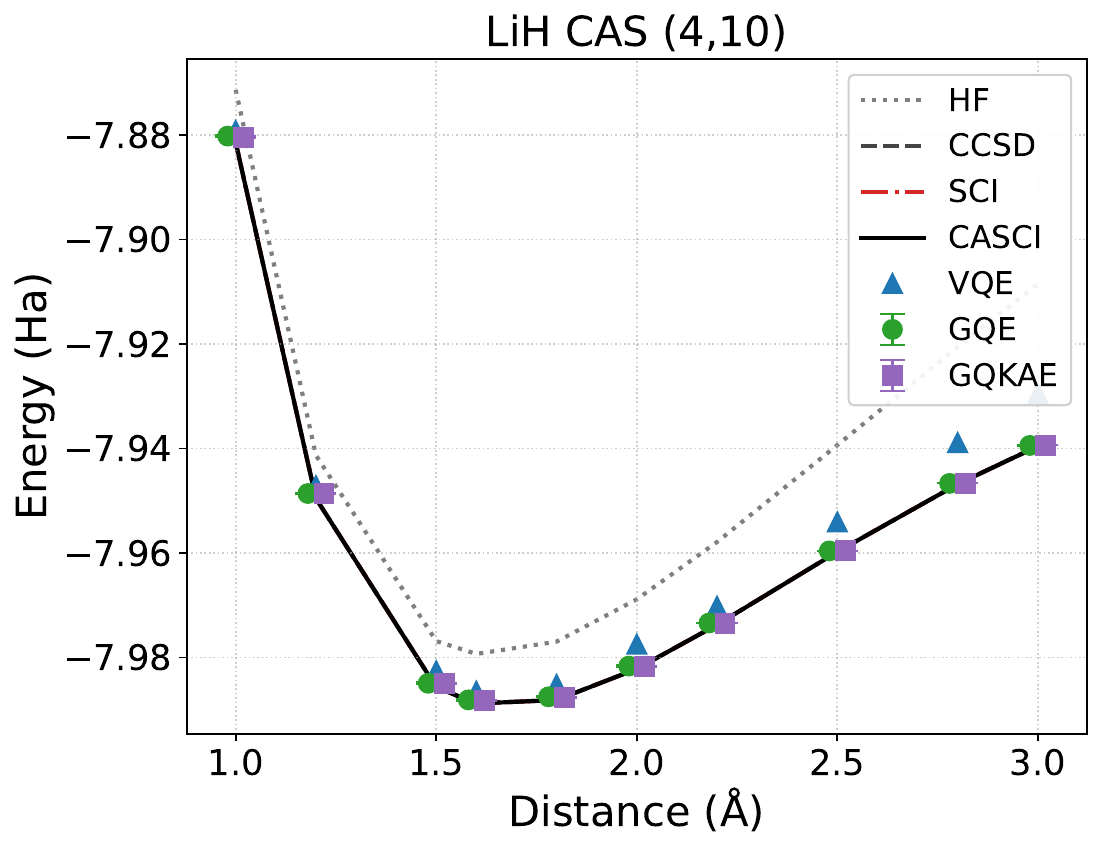}
        \caption{LiH (20 qubits)}
        \label{fig:energy_lih}
    \end{subfigure}

    \vspace{2pt}

    \begin{subfigure}{0.32\textwidth}
        \centering
        \includegraphics[width=\linewidth]{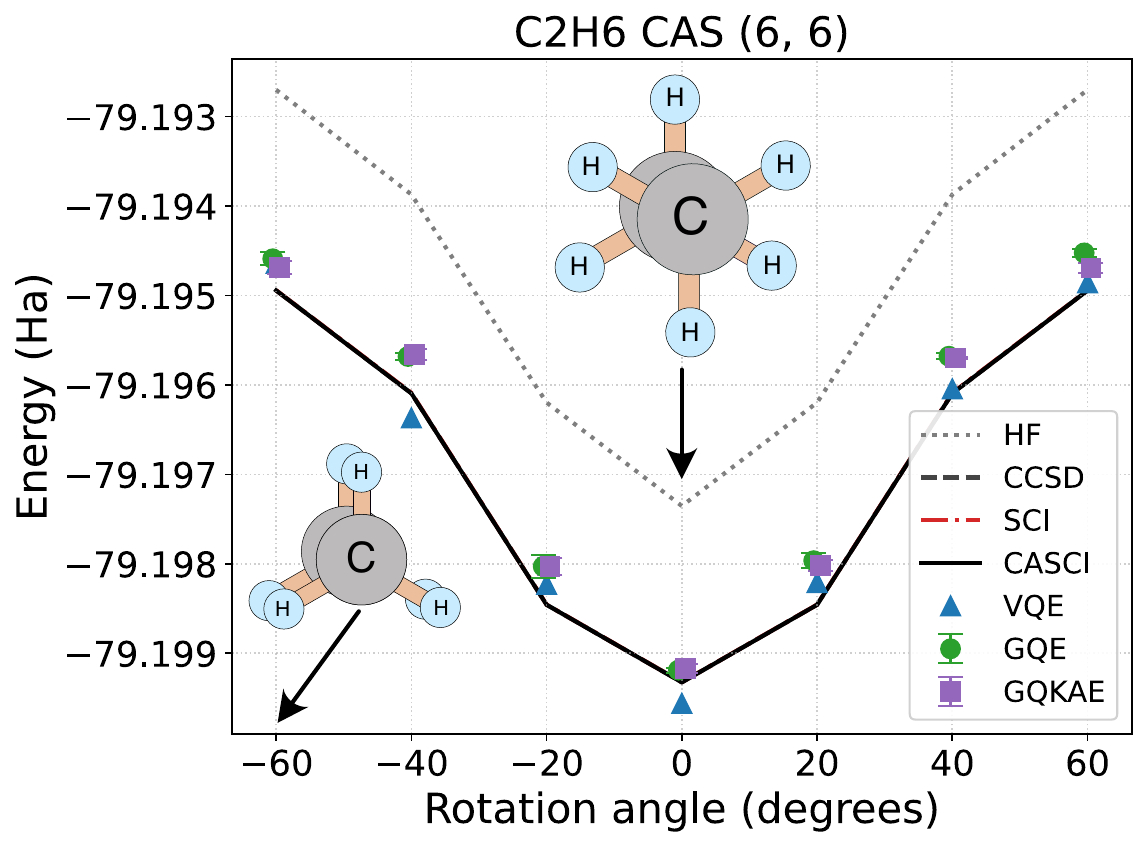}
        \caption{$\text{C}_2\text{H}_6$ (12 qubits)}
        \label{fig:energy_c2h6}
    \end{subfigure}
    \hfill
    \begin{subfigure}{0.32\textwidth}
        \centering
        \includegraphics[width=\linewidth]{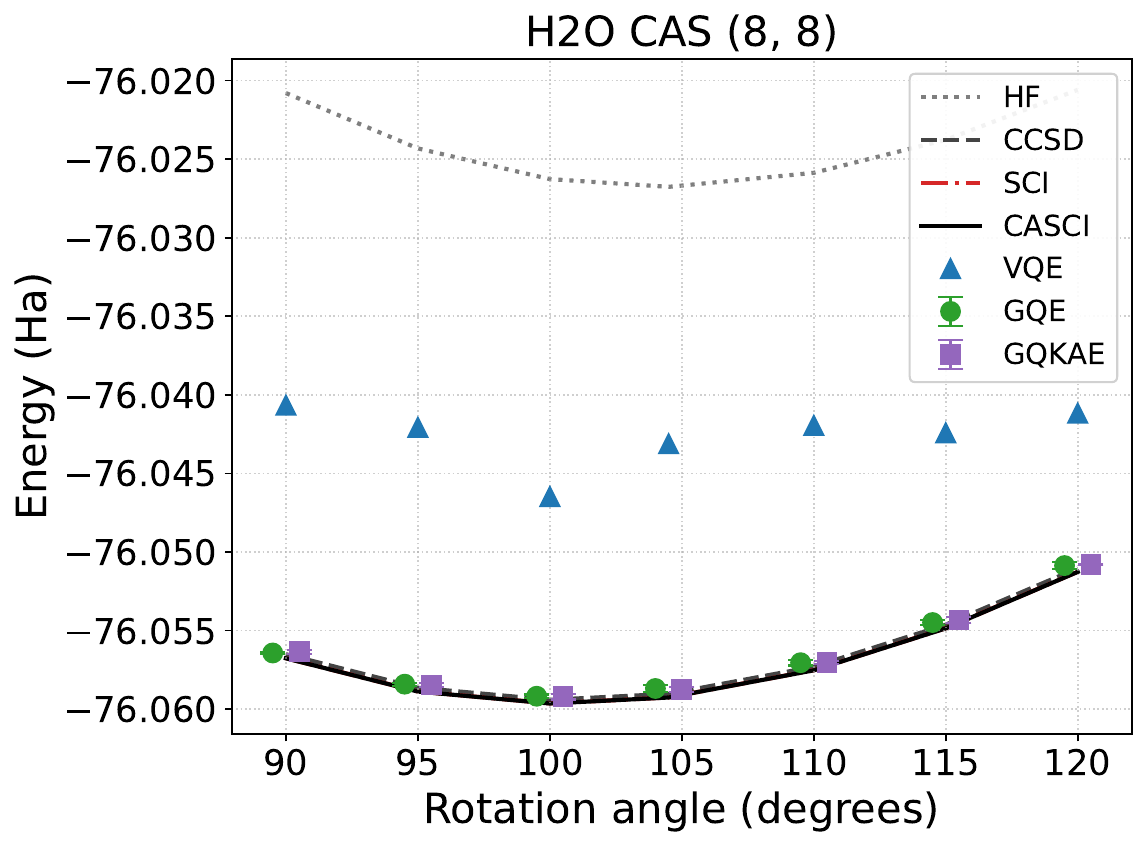}
        \caption{$\text{H}_2\text{O}$ (16 qubits)}
        \label{fig:energy_h2o}
    \end{subfigure}
    \hfill
    \begin{subfigure}{0.32\textwidth}
        \centering
        \includegraphics[width=\linewidth]{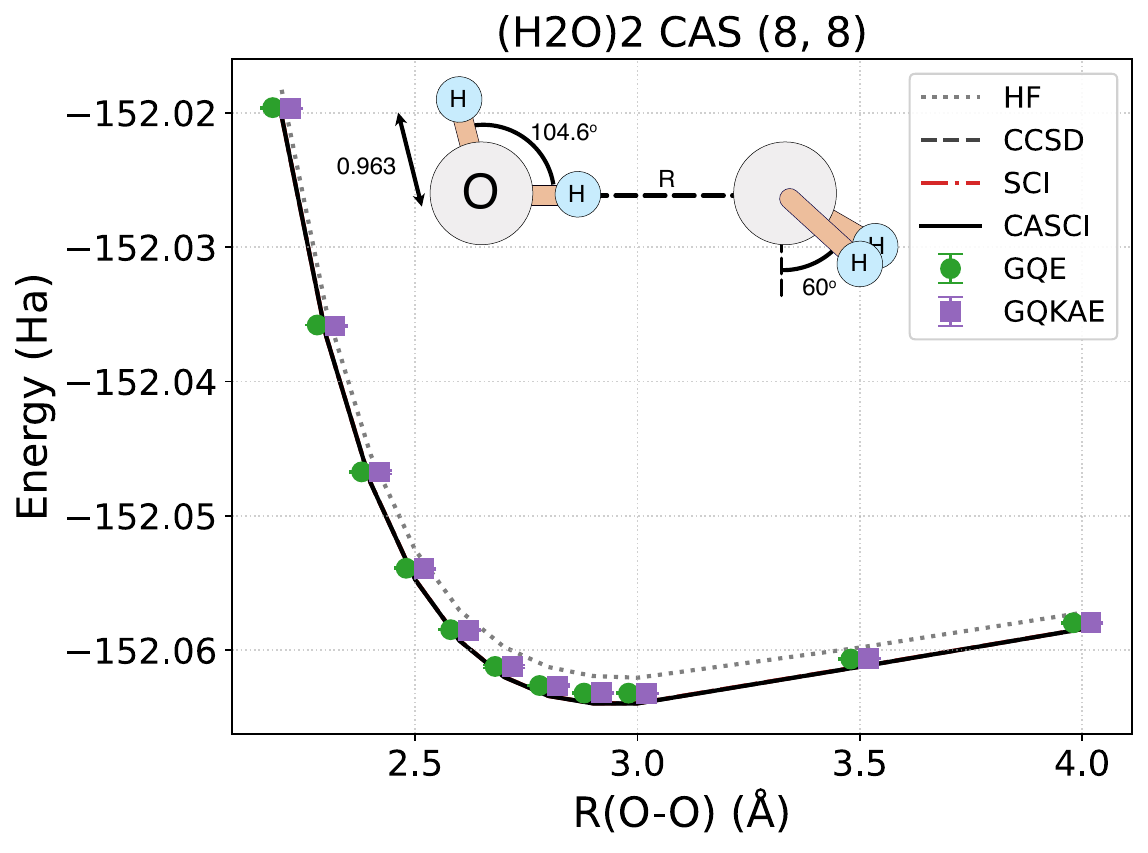}
        \caption{$\text{H}_2\text{O}$ dimer (16 qubits)}
        \label{fig:energy_h2o_dimer}
    \end{subfigure}
    \
    \caption{Potential energy surface of the six target molecules. Subfigures present the energy variations with respect to bond lengths, bond angles, and intermolecular distances. The predictions of GQKAE (purple squares) and GQE (green circles) are compared against HF (grey dotted lines), CCSD (grey dashed lines), SCI (red dash-dot lines), CASCI (black solid lines), and VQE (blue triangles). The sequence lengths ($L$) of the generated circuits were fixed based on computational efficiency tradeoffs: $L=20$ for H$_4$, $L=90$ for N$_2$, $L=70$ for LiH, $L=20$ for C$_2$H$_6$, $L=130$ for H$_2$O and $L=50$ for (H$_2$O)$_2$. }
    \label{fig:energy_all}
    \vspace{-10pt}
\end{figure*}

\begin{figure*}[t]
    \centering
    
    \begin{subfigure}{0.32\textwidth}
        \centering
        \includegraphics[width=\linewidth]{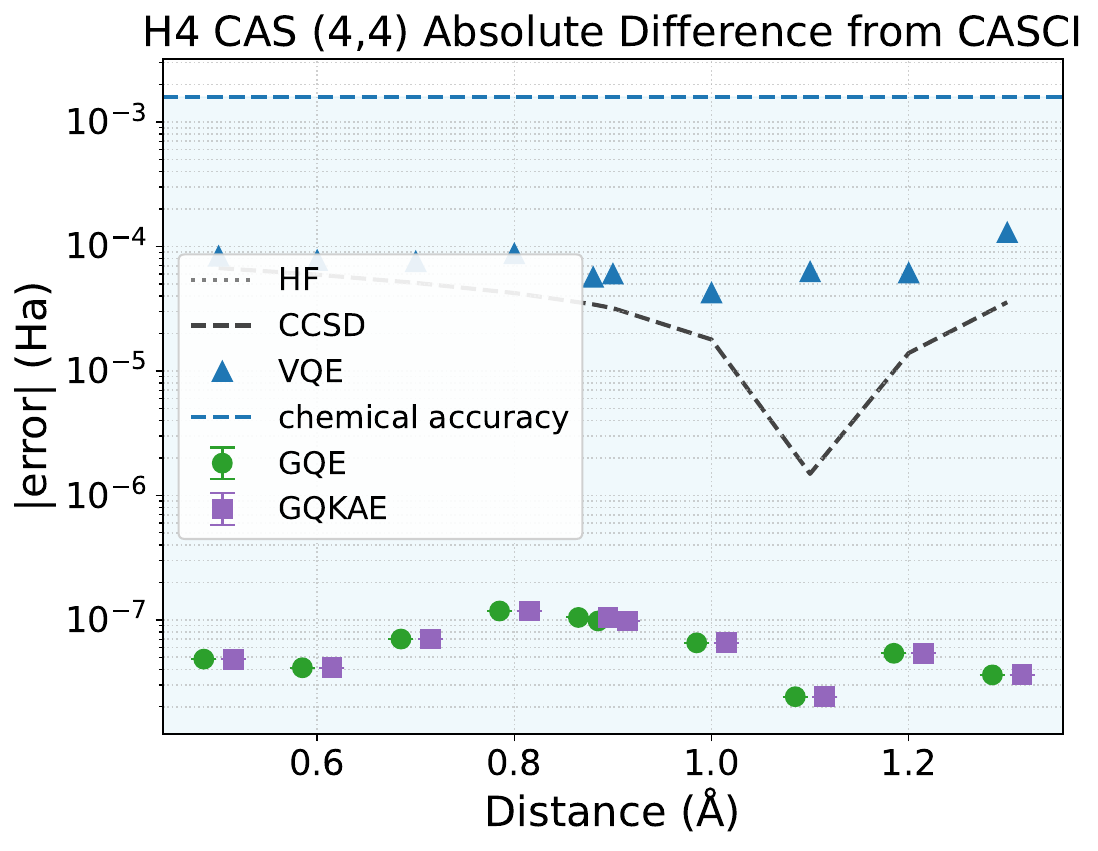}
        \caption{$\text{H}_4$ (8 qubits)}
        \label{fig:error_h4}
    \end{subfigure}
    \hfill
    \begin{subfigure}{0.32\textwidth}
        \centering
        \includegraphics[width=\linewidth]{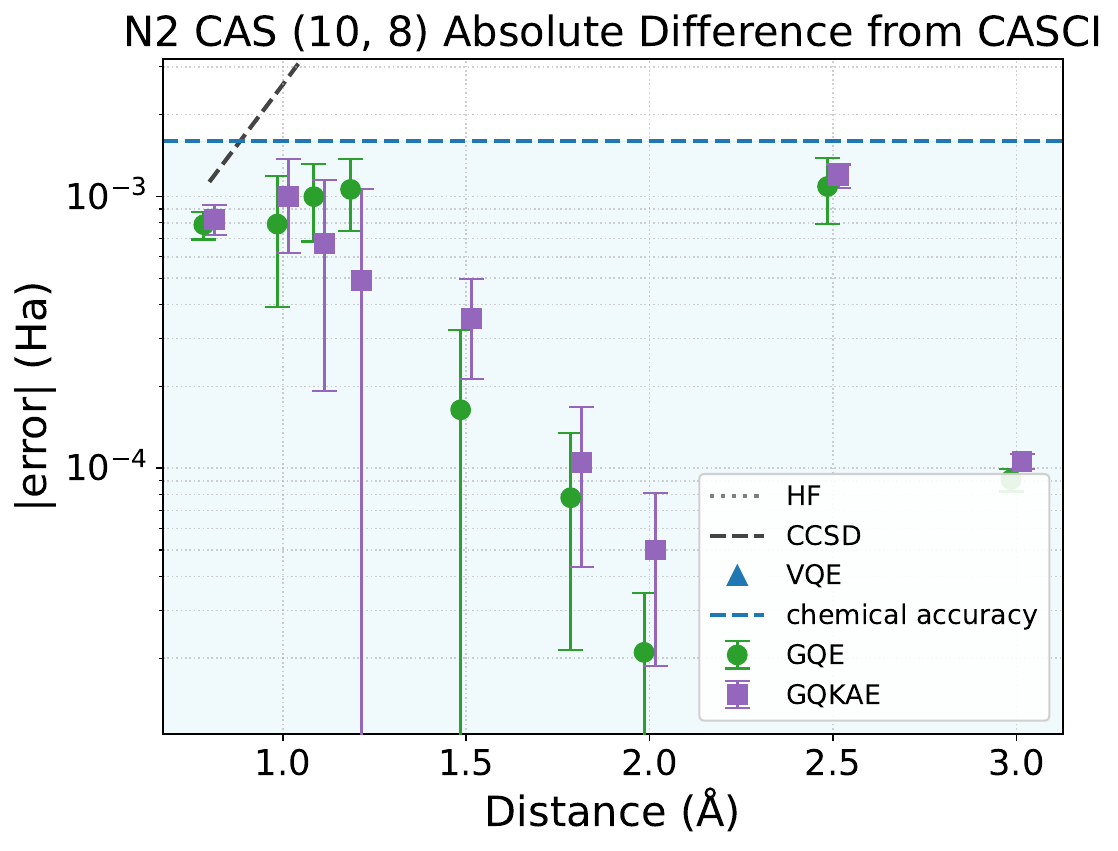}
        \caption{$\text{N}_2$ (16 qubits)}
        \label{fig:error_n2}
    \end{subfigure}
    \hfill
    \begin{subfigure}{0.32\textwidth}
        \centering
        \includegraphics[width=\linewidth]{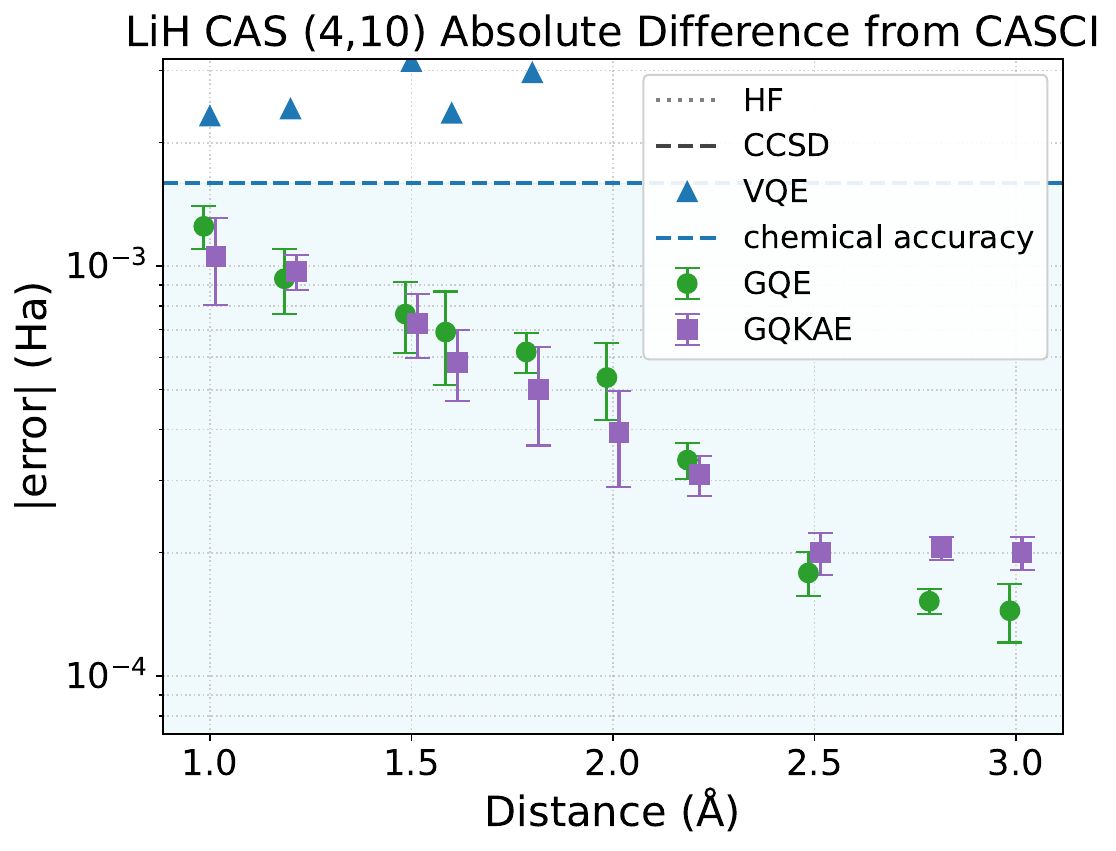}
        \caption{LiH (20 qubits)}
        \label{fig:error_lih}
    \end{subfigure}

    \vspace{2pt}

    \begin{subfigure}{0.32\textwidth}
        \centering
        \includegraphics[width=\linewidth]{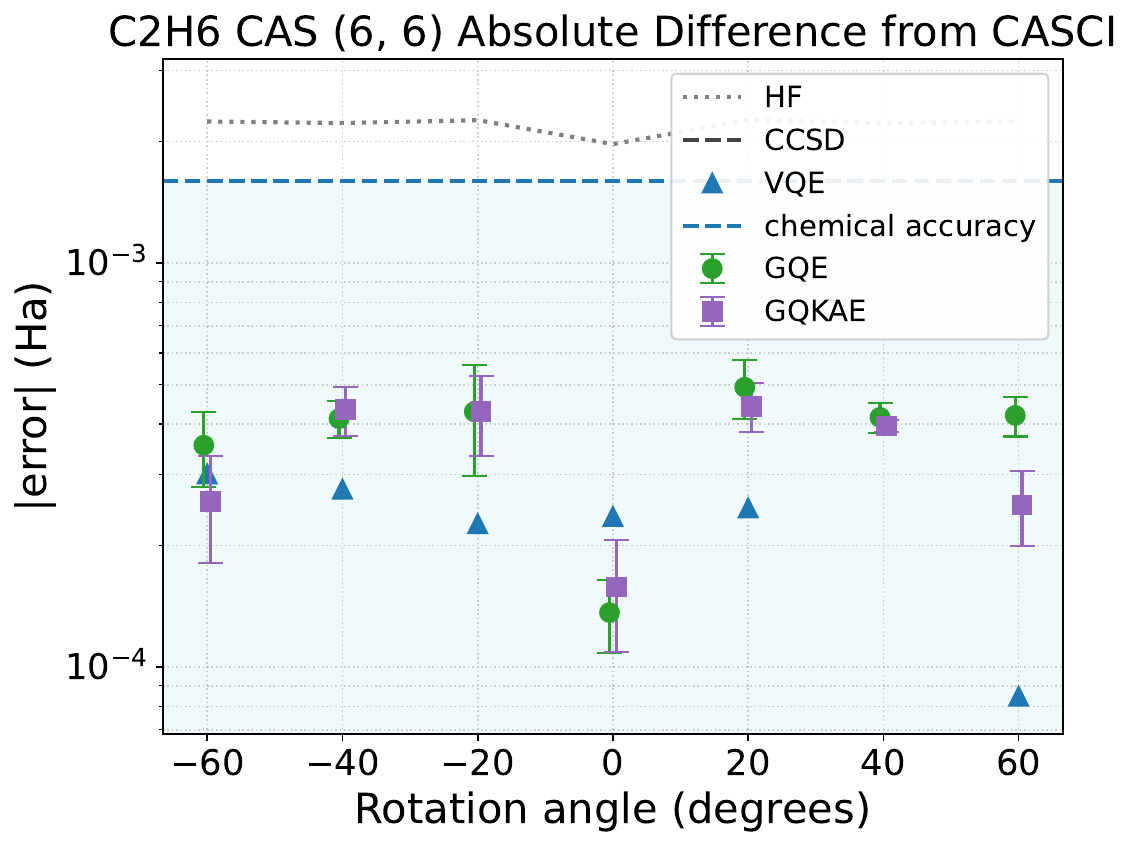}
        \caption{$\text{C}_2\text{H}_6$ (12 qubits)}
        \label{fig:error_c2h6}
    \end{subfigure}
    \hfill
    \begin{subfigure}{0.32\textwidth}
        \centering
        \includegraphics[width=\linewidth]{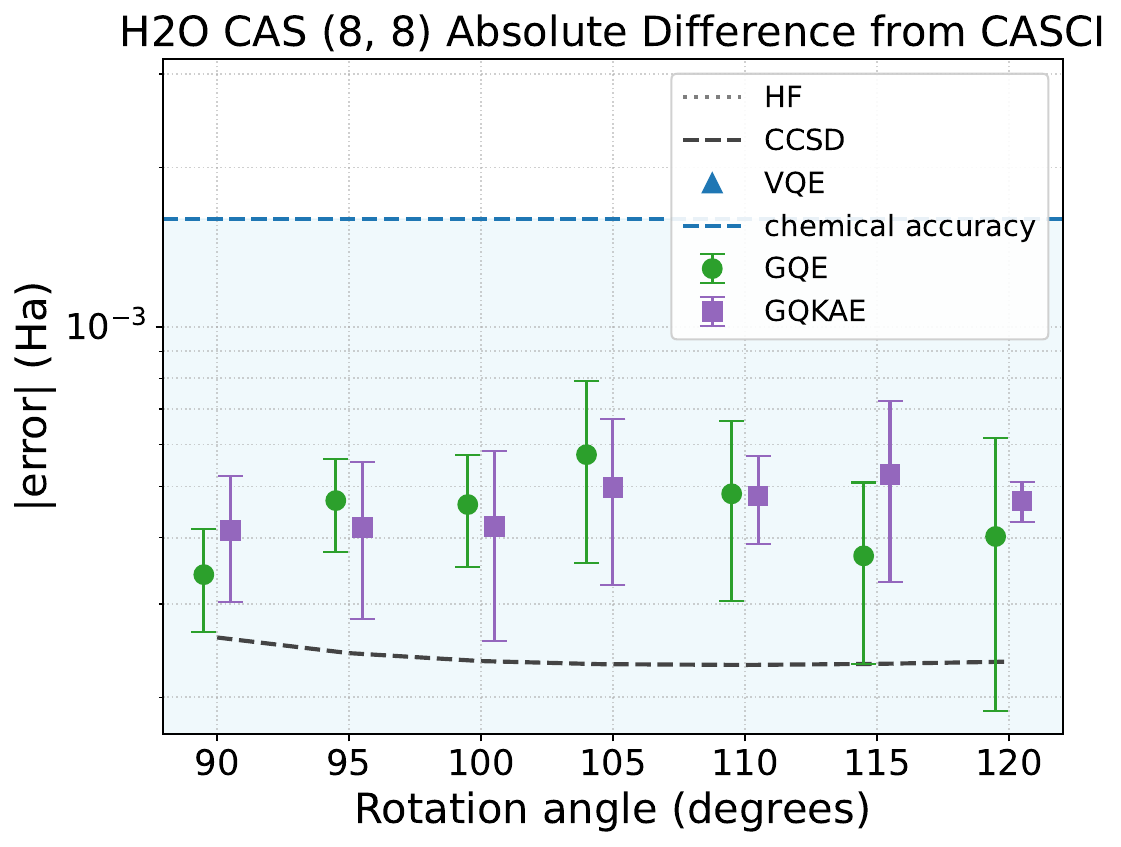}
        \caption{$\text{H}_2\text{O}$ (16 qubits)}
        \label{fig:error_h2o}
    \end{subfigure}
    \hfill
    \begin{subfigure}{0.32\textwidth}
        \centering
        \includegraphics[width=\linewidth]{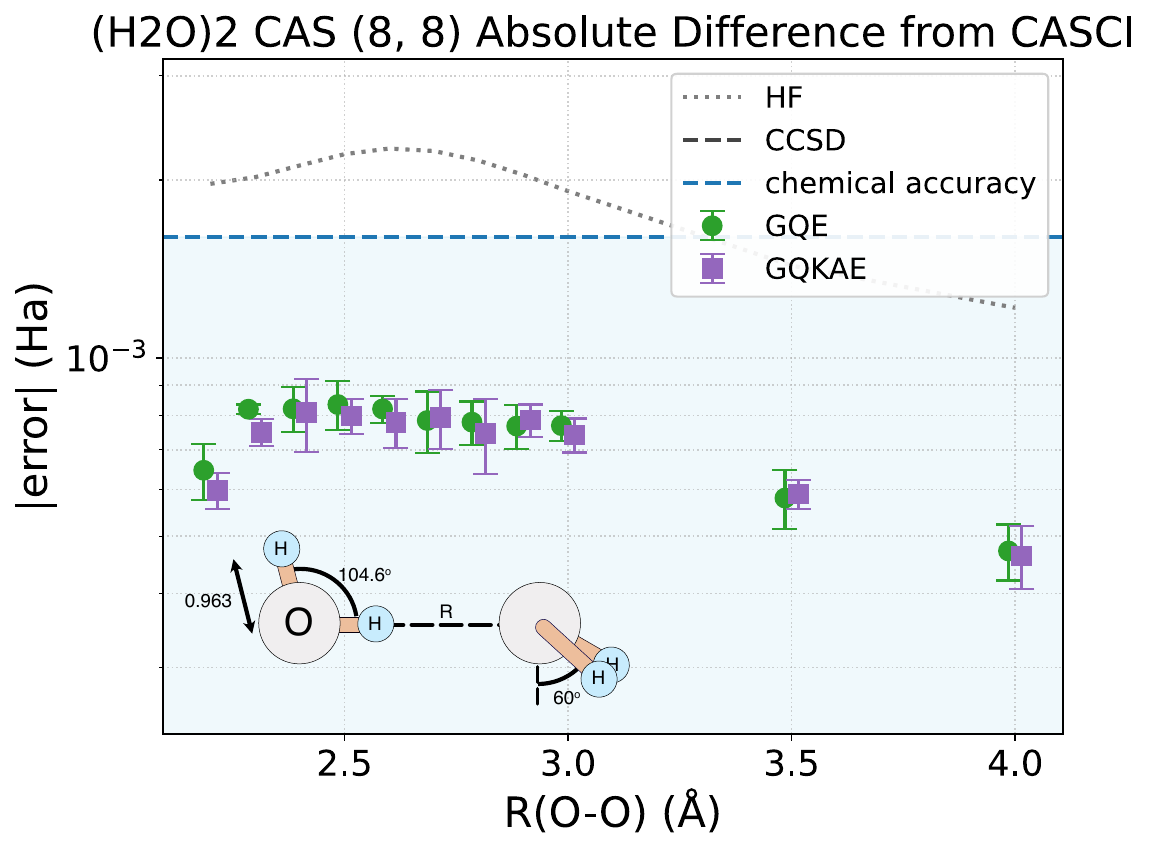}
        \caption{$\text{H}_2\text{O}$ dimer (16 qubits)}
        \label{fig:error_h2o_dimer}
    \end{subfigure}

    \caption{Absolute error from the CASCI energy on a logarithmic scale for the six molecular systems. The blue dashed line represents the chemical accuracy threshold. Reference lines for HF, CCSD, and VQE are included for comparative context. Error bars indicate one standard deviation over five independent trials. }
    \label{fig:error_all}
    \vspace{-10pt}
\end{figure*}

To comprehensively evaluate the physical reliability of the generated quantum circuits, we mapped the potential energy surfaces (PES) across the six molecular systems. Fig.~\ref{fig:energy_all} illustrates the minimum energies obtained by the GQKAE and GQE frameworks as functions of bond length, bond angle, or intermolecular distance. To establish a rigorous baseline, the generative models are compared against the HF, CCSD, SCI , CASCI, and VQE references. 

As shown in Fig.~\ref{fig:energy_all}, both generative models successfully capture the qualitative behavior of the bond dissociation curves for H$_4$, N$_2$, and LiH. In these regions, where the stretching of chemical bonds introduces strong static correlation, single-reference methods such as HF exhibit significant deviations. In contrast, the circuits generated by GQE and GQKAE track the exact CASCI energy profiles closely across the entire dissociation coordinates. This indicates that the discrete optimization over the operator pool successfully constructs highly entangled states capable of describing multi-reference character without relying on continuous parameter tuning.

Beyond bond stretching, we investigated conformational variations--specifically C$_2$H$_6$ internal rotation and H$_2$O angle bending, as depicted in Fig.~\ref{fig:energy_c2h6} and~\ref{fig:energy_h2o}. For the ethane (C$_2$H$_6$) molecule, the geometric coordinates were constructed such that the relative torsional angle of $0^\circ$ denotes the equilibrium staggered conformation. Consequently, internal rotations to $\pm 60^\circ$ correspond to the fully eclipsed conformations. The torsional barrier (the energy difference between the eclipsed and staggered conformations) is a critical benchmark of quantum chemistry methods. Our calculations show that GQKAE predicts a torsional barrier 0.122 eV (0.00448 Ha), while GQE yields  0.125 eV (0.004599 Ha). Both results are in excellent agreement with the experimental value of approximately 0.13 eV~\cite{Shang2023C2H6}. It is worth noting that for C$_2$H$_6$, the VQE energies appear marginally lower than the CCSD references at certain points; this is primarily attributed to minor discrepancies in active orbital ordering and frozen-core mapping between the classical (PySCF) and quantum (CUDA-Q) backend solvers, which does not affect the validity of the relative energetic trends.

Finally, for the angle bending in H$_2$O and the intermolecular distance variations in the water dimer, the adoption of the cc-pVDZ basis set allows the models to accurately capture delicate dynamic correlations. The results confirm that both GQKAE and GQE smoothly reproduce the potential wells, demonstrating that the generated discrete gate sequences possess sufficient flexibility to describe both significant orbital symmetry shifts and weak non-covalent intermolecular forces.

\subsection{Result of Absolute Error }

Fig.~\ref{fig:error_all} plots the absolute energy differences relative to CASCI on a logarithmic scale, demarcating the chemical accuracy threshold (1.6 mHa) alongside HF, CCSD, and VQE baselines. For H$_4$, C$_2$H$_6$, and LiH (Fig.~\ref{fig:error_h4},~\ref{fig:error_c2h6}, and~\ref{fig:error_lih}), both GQKAE and GQE exhibit outstanding stability and accuracy. Across all scanned distances and angles, the absolute errors remain firmly situated within or well below the chemical accuracy zone. 

In the case of the N$_2$ dissociation profile (Fig.~\ref{fig:error_n2}), we observe a distinct phenomenon. Between the bond lengths of 1.0 \r{A} and 2.0 \r{A}, the standard deviations for both models expand significantly. This region corresponds to the breaking of the nitrogen triple bond, a classical hallmark of strong static correlation. This region corresponds to the breaking of the nitrogen triple bond, a classical hallmark of strong static correlation. The emergence of nearly degenerate electronic configurations creates a highly complex optimization landscape; consequently, different random seeds can drive the policy network to explore alternative, yet energetically competitive, subspace compositions.

For H$_2$O and the water dimer, utilizing the cc-pVDZ basis set yields excellent accuracy. Unlike less expressive models that struggle in strongly overlapping regimes, the absolute errors for both GQKAE and GQE remain robustly below the chemical accuracy threshold across various bending angles and close intermolecular distances. This confirms the models' capacity to efficiently resolve high-order non-covalent interactions and dynamic correlations.

\subsection{Impact of Measurement Shots and Subspace Truncation}\label{sec:shot}
\begin{figure}[t]
\centering
\includegraphics[width=\columnwidth]{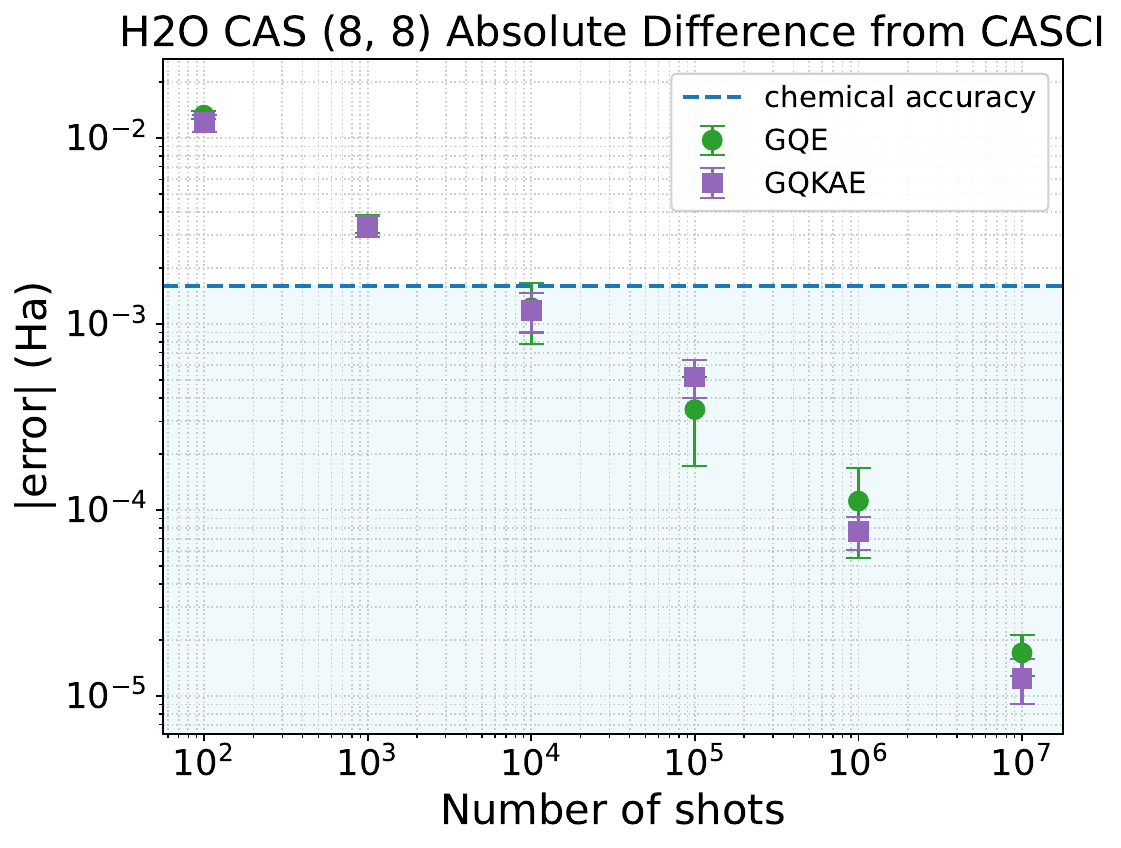}
\caption{Absolute energy error as a function of measurement shots for H$_2$O at its equilibrium angle, with fixed $d_{\text{max}}=2000$.
    }
\label{fig:shots}
\vspace{-10pt}
\end{figure}

\begin{figure}[t]
\centering
\includegraphics[width=\columnwidth]{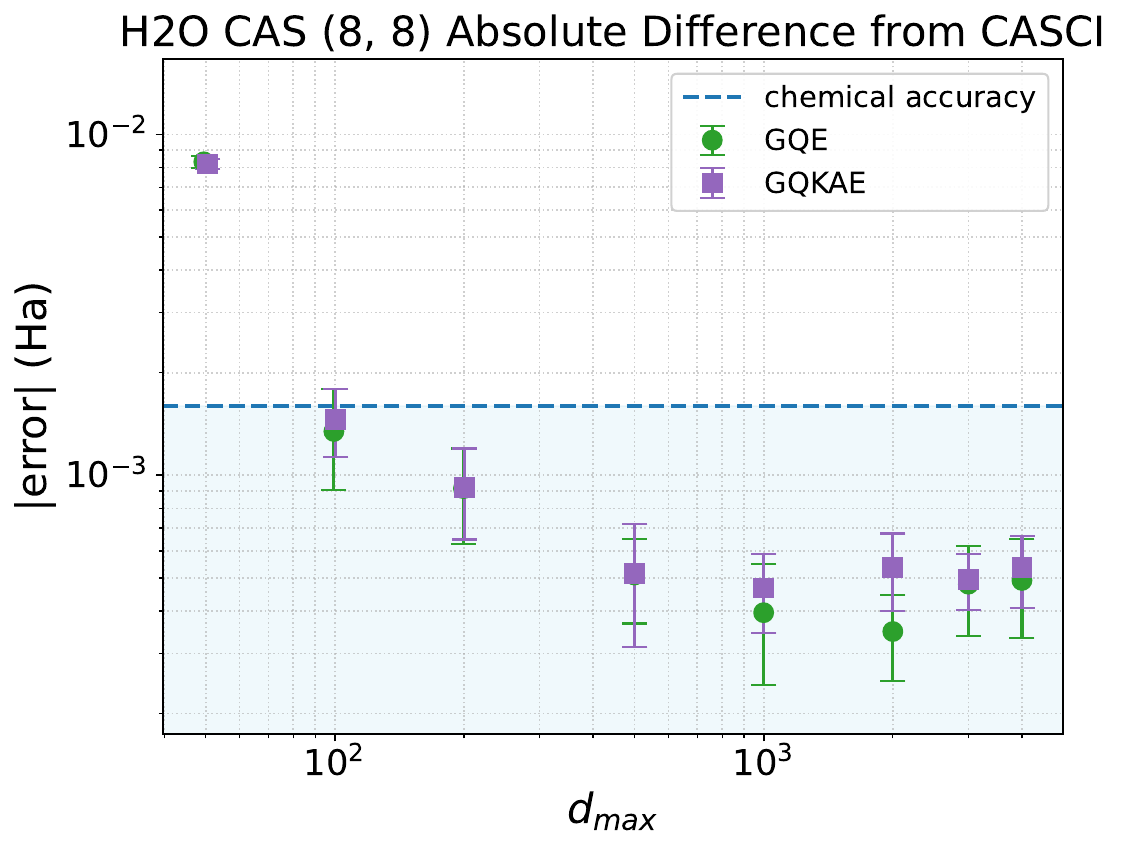}
\caption{Absolute energy error as a function of the $d_{\text{max}}$ for H$_2$O at its equilibrium angle, evaluated with a fixed number of measurement shots $N_\text{shots}=10^5$.
    }
\label{fig:dmax}
\vspace{-10pt}
\end{figure}

To evaluate robustness against statistical noise and classical post-processing constraints, we analyzed H$_2$O energy error at  equilibrium angle ($104.5^\circ$). Fig.~\ref{fig:shots} shows that with $d_{\text{max}}$ fixed at 2000, the error steadily decreases as measurement shots increase, crossing the chemical accuracy threshold at $10^4$ shots. Similarly, Fig.~\ref{fig:dmax} demonstrates that with $N_{\text{shots}}$ fixed at $10^5$, the error achieves chemical accuracy at $d_{\text{max}}=100$ and plateaus beyond 1000, indicating critical electronic configurations are captured. Notably,  GQKAE and standard GQE exhibit nearly identical trajectories in both analyses. This confirms that GQKAE's architectural compression does not compromise the sampled probability distribution, maintaining stable performance under limited resources.

\subsection{Quantum Resources and Parameter Efficiency}\label{sec:params}

\begin{table*}[b]
\centering
\caption{\textbf{Quantum gate counts for state preparation circuits.} Gate counts are evaluated assuming all-to-all connectivity and a standard decomposition into arbitrary-angle single-qubit rotation gates, CX gates, and single-qubit Clifford gates. Best result is shown in \textbf{bold}. }
\label{tab:gate_counts}
\footnotesize
\setlength{\tabcolsep}{8pt}
\begin{tabular}{lcc|rr|rr|rr}
\hline\hline
Mol. & Active sp. & Geometry & \multicolumn{2}{c|}{VQE} & \multicolumn{2}{c|}{GQE } & \multicolumn{2}{c}{GQKAE } \\
&            &          & 2-qubit & Total & 2-qubit & Total & 2-qubit & Total \\
\hline
H$_4$      & (4e, 4o)  & 0.88  & 1,312  & 3,108  & 100.4 $\pm$ 6.1 & 315.0 $\pm$ 16.0 & \textbf{100.0 $\pm$ 3.7} & \textbf{314.0 $\pm$ 15.0} \\
N$_2$      & (10e, 8o) & 1.1   & 31,760 & 70,550 & 518.0 $\pm$ 19.3 & 1,603.4 $\pm$ 32.0 & \textbf{506.8 $\pm$ 13.1} & \textbf{1,559.4 $\pm$ 37.4} \\
LiH        & (4e, 10o) & 1.6   & 41,600 & 90,884 & 396.8 $\pm$ 8.2 & 1,220.4 $\pm$ 22.0 & \textbf{391.6 $\pm$ 14.3} & \textbf{1,212.6 $\pm$ 30.4} \\
C$_2$H$_6$ & (6e, 6o)  & 0.0   & 9,024  & 20,538 & \textbf{114.0 $\pm$ 4.5} & \textbf{339.0 $\pm$ 5.4}  & 115.2 $\pm$ 3.4 & 347.2 $\pm$ 9.1 \\
H$_2$O     & (8e, 8o)  & 104.5 & 36,480 & 81,032 & \textbf{737.20 $\pm$ 16.71} & \textbf{2284.60 $\pm$ 54.72} & 747.60 $\pm$ 5.55 & 2334.20 $\pm$ 22.95 \\
(H$_2$O)$_2$ & (8e, 8o) & 2.9   & 36,480 & 81,032 & 289.60 $\pm$ 6.07 & 909.20 $\pm$ 14.81 & \textbf{286.40 $\pm$ 2.19} & \textbf{878.80 $\pm$ 20.08} \\
\hline\hline
\end{tabular}
\vspace{-5pt}
\end{table*}

\begin{table*}[t] 
\centering 
\caption{Comparison of trainable classical parameters, the corresponding parameter memory, and the wall time for GQE and GQKAE. GQKAE achieves an approximately \textbf{66\% reduction} in both parameter count and parameter memory relative to GQE, while using $\sim10\%$ less wall time overall.} 
\label{tab:model_parameters} 
\footnotesize 
\resizebox{\textwidth}{!}{ 
\begin{tabular}{l|ccc|ccc|c} 
\hline\hline 
& \multicolumn{3}{c|}{GQE} & \multicolumn{3}{c|}{GQKAE} \\ 
Mol. & \# Params (M) & Memory (MB) & Wall time (s)& \# Params (M) & Memory (MB) & Wall time (s) & Speedup (\%)\\ 
\hline 
H$_4$ & 55.26 & 162.90 & 216.6 & 14.3 & 42.6 & 198.1 & 13.6 \\ 
N$_2$ & 55.24 & 162.88 & 326.7 & 14.5 & 42.7 & 292.8 & 9.0 \\ 
LiH & 55.24 & 162.88 & 283.3 & 14.5 & 42.7 & 263.1 & 9.8 \\ 
C$_2$H$_6$ & 54.76 & 162.41 & 218.8 & 14.4 & 42.6 & 200.1 & 16.7 \\ 
H$_2$O & 42.8 & 163.12 & 254.8 & 14.6 & 55.48 & 242.6 & 6.8 \\ 
(H$_2$O)$_2$ & 42.8 & 163.27 & 247.2 & 14.6 & 55.62 & 197.0 & 13.3 \\ 
\hline\hline 
\end{tabular} 
} 
\vspace{-13pt} 
\end{table*}

While sampling efficiency and energy accuracy are critical, the practical implementability of QSCI on near-term quantum hardware hinges fundamentally on circuit resources costs. Table~\ref{tab:gate_counts} summarizes total and two-qubit gate counts for VQE, GQE, and GQKAE. Table~\ref{tab:gate_counts} shows both GQKAE and GQE exhibit comparable gate requirements, achieving orders-of-magnitude reductions over the UCCSD-VQE baseline. While advanced adaptive strategies (e.g. ADAPT-VQE) can also reduce circuit depth, we employ the standard uncompiled UCCSD-VQE as a chemically motivated baseline to underscore the massive resource savings achieved strictly through our generative discrete search. For instance, in the H$_2$O system, VQE necessitates over 81,000 total gates and 36,480 two-qubit gates. In stark contrast, GQKAE reaches comparable or superior accuracy using only 2,334 total gates and 747 two-qubit gates. This massive reduction in circuit depth mitigates the accumulation of hardware noise.

Beyond the quantum gate reductions, the true architectural distinction lies in the classical computational overhead. Table~\ref{tab:model_parameters} presents the total number of trainable parameters memory footprint and wall time for both generative models. By integrating the HQKAN module into the highly compressed latent space, the parameter count drops by approximately 66$\%$--from roughly 42.7 M to just 14.5 M. 

Crucially, this algorithmic compression yields practical computational benefits: runtime memory drops from over 162 MB in standard GQE to $\sim$55 MB in GQKAE. Furthermore, GQKAE achieves a noticeable reduction in circuit generation wall time, yielding speedups ranging from 6.8$\%$ to 16.7$\%$ across the tested molecules. This highlights that the QKAN-based architecture not only minimizes the classical hardware footprint but also actively accelerates the inference process, providing a highly scalable foundation for hybrid HPC-quantum co-design.

\section{Conclusion}\label{sec:Discussion}

GQKAE successfully captures complex chemical phenomena--including delicate dynamic correlations and weak intermolecular forces--while reducing trainable parameters and memory by ~66$\%$ relative to the GPT-2 baseline. Furthermore, it demonstrates robust performance under limited measurement shots and subspace truncation constraints, thereby achieving chemical accuracy, computational speedups, and parameter reduction simultaneously, contributing a highly efficient advancement to the broader GQE framework. Currently, the circuit sequence length ($L$) must be manually specified. Future work could integrate Graph Neural Networks~\cite{minami2025generative}  into the encoder to autonomously determine $L$ and enable zero-shot transfer learning across molecules.

While the current evaluations were performed using classical simulators, the practical advantage of GQKAE lies in its ability to circumvent trainability issues such as barren plateaus~\cite{wang2021noise,Larocca2025barrenplateaus,mcclean2018barren} by replacing parameterized quantum circuits with a classical generative model, while simultaneously generating significantly shallower circuits than standard VQE. This reduction in both optimization difficulty and gate counts makes the generated circuits significantly more viable for near-term devices and early FTQC (Fault-Tolerant Quantum Computing) architectures. A critical next step is to deploy these optimized circuits on real quantum hardware to investigate their performance under actual device noise. Combining our framework with advanced error mitigation techniques will be essential to fully realize quantum utility in practical chemistry problems.

GQKAE suggests a paradigm where compressed generative classical models, rather than fixed parameterized circuits, drive quantum-state preparation. By drastically reducing the classical-side memory and runtime overhead of circuit generation, the QKAN-based backbone provides a highly scalable route for HPC--quantum co-design, where generative circuit models can be pretrained across families of Hamiltonians and fine-tuned for target systems.

\section*{Acknowledgment}
The authors express their gratitude to Kohei Nakaji and Jin-Sung Kim from NVIDIA Quantum team for their invaluable insights and comments, which were instrumental in the success of this research.
Y.-C. Hsu, J.-C. Jiang, C.-H. Lin and K.-C. Peng thank the National Center for High-Performance Computing (NCHC), National Institutes of Applied Research (NIAR), Taiwan, for providing computational and storage resources supported by the National Science and Technology Council (NSTC), Taiwan, under Grants No. NSTC 114-2119-M-007-013.
The authors thank NVIDIA AI Technology Center (NVAITC) for their technical support.

\clearpage
\bibliographystyle{IEEEtran}
\bibliography{reference}

\end{document}